\documentstyle[12pt,aaspp4]{article}

\begin{document}

\title{WHAM Observations of H$\alpha$ Emission from High Velocity Clouds 
in the M, A, and C Complexes}

\author{S. L. Tufte, R. J. Reynolds, \& L. M. Haffner}
\affil{Astronomy Department, University of Wisconsin -- Madison}
\authoraddr{475 North Charter Street, Madison, WI  53706}

\begin{abstract}

The first observations of the recently completed Wisconsin H-Alpha
Mapper (WHAM) facility include a study of emission lines from high
velocity clouds in the M, A, and C complexes, with most of the
observations on the M~I cloud.  We present results including clear
detections of H$\alpha$ emission from all three complexes with
intensities ranging from 0.06 to 0.20 R.  In every observed direction
where there is significant high velocity H~I gas seen in 21 cm
emission we have found associated ionized hydrogen emitting the
H$\alpha$ line.  The velocities of the H$\alpha$ and the 21 cm
emissions are well correlated in every case except one, but the
intensities are not correlated.  There is some evidence that the
ionized gas producing the H$\alpha$ emission envelopes the 21 cm
emitting neutral gas but the H$\alpha$ ``halo'', if present, is not
large.  If the H$\alpha$ emission arises from the photoionization of
the H~I clouds, then the implied incident Lyman continuum flux
F$_{LC}$ at the location of the clouds ranges from 1.3 to 4.2 $\times$
10$^{5}$ photons cm$^{-2}$ s$^{-1}$.  If, on the other hand, the
ionization is due to a shock arising from the collision of the
high-velocity gas with an ambient medium in the halo, then the density
of the pre-shocked gas can be constrained.  We have also detected the
[S~II] $\lambda$6716 line from the M~I cloud and have evidence that
the [S~II] $\lambda$6716 to H$\alpha$ ratio varies with location on
the cloud.

\end{abstract}


\section{Introduction}
High velocity clouds (HVCs) were discovered over 30 years ago through
the 21 cm line of neutral hydrogen (Muller, Oort, and Raimond
\markcite{Muller63}1963).  While they have been studied extensively
(and almost exclusively) in the 21 cm line since then, the origin and
nature of HVCs is still largely unknown.  They are observationally
defined as gas moving at velocities not explainable by cylindrical
galactic rotation (a rule of thumb is $|v_{lsr}| \geq$ 90 km
s$^{-1}$).  In the northern hemisphere, the vast majority of the HVCs
have negative velocities, that is, they are falling toward the
Galactic plane.  In addition, there exists gas at intermediate
velocities (IVCs), also primarily falling toward the plane in the
northern hemisphere, which may have a connection with the HVCs (see
Kuntz \& Danly \markcite{Kuntz96}1996).  Many theories have been
proposed to explain the HVCs.  One model suggests that they are
remnants of the formation of the Milky Way only now being captured by
the gravitational field of the Galaxy (Oort \markcite{Oort66}1966,
Oort \markcite{Oort70}1970, Wakker \markcite{Wakker90}1990).  Another
suggests that they are the cooled condensate of material vented into
the halo as a result of super-bubble breakout (the galactic fountain
model, see Shapiro \& Field \markcite{Shapiro76}1976, Bregman
\markcite{Bregman80}1980).  Another proposal (Murai \& Fujimoto
\markcite{Murai80}1980, Giovanelli \markcite{Giovanelli81}1981) is
that the gas was tidally stripped from the Magellanic clouds as they
orbit the galaxy.  There is convincing evidence that this is the case
for a southern complex of high velocity gas known of as the Magellanic
Stream, but it is harder to associate the northern HVCs with this
origin.  For a comprehensive review of HVCs, see Wakker \& Van Woerden
\markcite{Wakker97}(1997).

One of the principal limitations to progress in understanding the HVCs
is the lack of knowledge of the distances to the emitting gas.  The
only reliable method of determining distances to HVCs is through
absorption line spectroscopy toward bright objects of known distance
in the direction of the cloud.  This method has now been successfully
employed to constrain the distance to the M complex and the A complex.
Toward the M complex, Danly et al. \markcite{Danly93}(1993) have
constrained the height z above the Galactic plane to 1.5 to 4.4 kpc.
Wakker et al. \markcite{Wakker96}(1996) and van Woerden et
al. \markcite{vanWoerden97}(1997) have constrained the A cloud to be
between 3 and 7 kpc above the plane.  This places these HVC complexes
in the Galactic halo and thus makes them excellent ``test particles''
for studying the conditions there.  The distance to the C complex is
still unknown.

While considerable effort has been expended looking for emission lines
from HVCs in the past, there have only been a few detections.  A
search for H$\alpha$ by Reynolds \markcite{Reynolds87}(1987) in 6
directions toward HVCs, including two directions toward the M complex,
yielded negative results.  The upper limits placed on the intensity by
this study ranged from 0.2 R to 0.6 R.  One rayleigh (R) is 
10$^{6} / 4\pi$ photons cm$^{-2}$ s$^{-1}$ sr$^{-1}$.  Kutyrev and Reynolds
\markcite{Kutyrev89}(1989) obtained a 4 $\sigma$ detection of a very
high velocity cloud (v$_{lsr}$ = -300 km~s$^{-1}$) in Cetus.  The
H$\alpha$ brightness in this direction was measured to be I$_{\alpha}$
= 0.08 R\@.  M\"{u}nch \& Pitz \markcite{Munch89}(1989) marginally
detected high velocity H$\alpha$ emission from the M~II cloud with
I$_{\alpha}$ = 0.15 R (one of our observation directions is very near
theirs, and we compare the results below).  Songaila et al.\
\markcite{Songaila89}(1989) claim a detection of H$\alpha$ from high
velocity gas in complex C\@.  The intensity they measured was 0.03 R;
however, we here present observations of the same direction and find
the emission to be significantly brighter than the Songaila \& Cowie
measurement.  Very recently, Weiner \& Williams
\markcite{Weiner96}(1996) have detected H$\alpha$ emission from a
number of directions toward the Magellanic Stream.  They find that
H$\alpha$ emission enhancements (typically $\simeq$ 0.2 R) are
associated with gradients in the H~I distribution, which they
interpret as evidence that the emission is produced by ram pressure
heating of the cloud as it passes through a low density ambient
medium.  In general, the previous observations of H$\alpha$ from HVCs
required long integration times and, with the exception of the
Magellanic Stream detections, produced null or marginal results.

Much can be learned about both the HVCs and the Galactic halo
environment from a study of optical emission lines.  For example, the
brightness of the H$\alpha$ line gives information on the amount of
ionized hydrogen that is associated with the neutral gas.  If the
ionizations are due to photoionization, then the H$\alpha$ intensity
is a direct measure of the Lyman continuum flux incident upon the
cloud.  Measurements of this flux at various locations above the plane
of the Galaxy can have an important bearing on more general questions
regarding ionized gas in the disk and halo of the Milky Way.  If the
ionizations are due to collisional processes induced by the rapid
motion of the cloud (i.e.  shocks), then the intensity is a measure of
the ambient density of material through which the cloud travels.  It
may be possible to distinguish between these two ionization mechanisms
by measuring other emission lines such as [O~III] $\lambda$5007,
[S~II] $\lambda$6716, and [N~II] $\lambda$6584.  It is also possible
that the pattern of H$\alpha$ intensity may shed some light on the
relationship between the neutral and ionized gas and perhaps features
in the ISM at lower velocity.

\section{Observations}
The Wisconsin H-Alpha Mapper (WHAM) is a recently completed facility
that is located at Kitt Peak, AZ and is operated remotely from
Madison, WI (Reynolds et al. \markcite{Reynolds98}1998; Tufte
\markcite{Tufte97}1997).  The WHAM instrument is optimized for the
detection and study at high spectral resolution of faint optical
emission lines from spatially diffuse astronomical sources.  By
employing the largest available Fabry-Perot etalons and a state of the
art CCD chip as an imaging detector, the WHAM instrument achieves
unprecedented sensitivity for velocity resolved studies of faint
emisson lines from diffuse sources.  This opens a new window to the
HVCs.

The new WHAM spectrometer is well suited to the study of emission
lines from HVCs for several reasons: 1) the Fabry-Perot spectrometer
is ideal for detecting spatially extended emission from faint sources
at high resolving power due to the much higher throughput available at
a given spectral resolution compared to grating spectrometers, 2) the
15 cm etalons used by the WHAM spectrometer are the largest available,
3) the 1\arcdeg\ beam is well matched to the large angular sizes
characteristic of many of the HVCs, and 4) the 12 km~s$^{-1}$
resolution is well matched to the 20--25 km~s$^{-1}$ (FWHM) line
widths (21 cm) typical of HVCs.  As a result, the WHAM spectrometer
allows us to obtain clear detections of emissions with intensity of
order 0.1 R in 30 minutes integration time.  

We obtained H$\alpha$ observations in 15 directions in the vicinity of
the high-velocity clouds M~I and M~II.  Table~\ref{tab4-obs_info}
lists the coordinates of each look direction along with the total
exposure time and the date(s) of observation.  Figure~\ref{fig4-map_m}
shows our look directions on a map of 21 cm contours from the
Leiden/Dwingeloo survey (Hartmann \& Burton
\markcite{Hartmann97}1997).  The 21 cm data measure the amount of
neutral hydrogen at high velocity (v$_{lsr}$ $\leq$ -80 km~s$^{-1}$).
The M complex H$\alpha$ study was conceived as a set of 8 ``on--off''
pairs.  The directions 1a, 2a, 5a, 6a, 7a, and 8a are ``on'' the M~I
cloud as defined by the 21 cm maps, and the associated ``off''
directions, which are 4a, 3a, 9a, 12a, 11a, and 10a, were chosen based
on their lack of detected H~I gas at high velocity. Note that
directions 9a and 11a do not appear on this map and that 7a, while
near the M~I cloud, is not within the the lowest 21 cm column density
contour.  For the M~II cloud, the ``on'' directions are 1b and 2b, and
the ``off'' direction is 3b (see Fig.~\ref{fig4-map_m}).  The M clouds
are especially interesting because of their high Galactic latitude
(50\arcdeg $\leq b \leq$ 70\arcdeg).  At this latitude, Galactic
rotation cannot be a significant factor in the radial velocity, which
is then a fairly good measure of motion perpendicular to the Galactic
plane.  Further, dust obscuration is less significant than at lower
latitudes.

We have also obtained H$\alpha$ observations in 4 directions toward
the A complex (2 ``on'' directions and 2 ``off'' directions) and 2
directions toward the C complex (1 ``on'' direction and 1 ``off''
direction).  The locations of these beams for the A complex and the C
complex are shown on maps of the 21 cm intensity contours in
Figure~\ref{fig4-map_a} and Figure~\ref{fig4-map_c}, respectively.
The ``off'' directions for the C complex are off the map.

We use the ``on'' minus ``off'' method in order to isolate the
emission due to the HVCs under study.  This requires an observation
``on'' the direction of interest, immediately followed by an
observation of a nearby direction that is ``off'' the high velocity
cloud.  This very powerful technique allows for the subtraction of the
sky spectrum and cancellation of many systematic uncertainties.  Since
the intermediate velocity gas (IVC) extends over the entire region
around complex M and is fairly smooth (at least in the 21 cm maps),
this method also serves to subtract off the IVC component, allowing
for the isolation of the HVC component.  Figure~\ref{fig4-hvc_on_off}
shows an example of the method.  The upper panel depicts the spectrum
of an ``on'' direction (derived from a 900 second exposure of the 1a
direction) as a solid line and the spectrum of an ``off'' direction
(derived from a 900 second exposure of the 4a direction taken
immediately after the 1a exposure) as a dashed line.  Both spectra
show the relatively bright geocoronal H$\alpha$ emission line on the
right and very little else.  The geocoronal emission line is placed
near the red edge of the bandpass for use in the velocity calibration.
The irregularity near the blue edge of the spectrum is a remnant of
the flat-fielding process.  The lower panel shows the result of
subtracting the ``off'' spectrum from the ``on'' spectrum.  Here one
can clearly see an emission component near -110 km~s$^{-1}$ due to the
M~I cloud.  On the right, the difference spectrum takes a steep dive
due to a combination of the fact that the geocoronal line is brighter
in the ``off'' direction and that the 4a direction has more emission
from lower velocity gas (see below).  In general, problems with
accurately correcting for the roll off at the edge of the aperture
make it difficult to measure extremely faint emission lines at
velocities between -200 km~s$^{-1}$ and -175 km~s$^{-1}$.  Problems
with accurately removing the geocoronal line make it difficult to
extract information at velocities greater than about -40 km~s$^{-1}$.
This leaves a velocity interval of 130 km~s$^{-1}$ within which we can
make good measurements of very faint emissions from high velocity gas.

We discovered several emission lines in the night sky spectrum that
are probably extremely weak, not previously detected OH lines.
Figure~\ref{fig4-oh_spect} shows the average of two spectra from the
February 24 observations.  On this night, the sky lines were the
brightest, although the same lines are present to some degree in all
of the spectra.  The evidence for the lines originating in the
atmosphere and not the Galaxy is: 1) the lines are present in all of
the spectra (both ``on'' and ``off''), 2) their brightness seems to
depend more on the night than the look direction, 3) the location of
the lines is fixed with respect to the geocoronal line and not fixed
with respect to the LSR, and 4) the lines appear to be very narrow, as
would be expected for atmospheric lines but not Galactic emission.
The night sky features were fitted with Gaussian components at
locations shown by the arrows in Figure~\ref{fig4-oh_spect}.  A
terrestrial spectrum was then subtracted from each spectrum as part of
the data reduction process.  This spectrum was formed using the
spectrum shown in Figure~\ref{fig4-oh_spect} as a template.  The
locations and relative strengths of the sky lines were held fixed and
a single parameter corresponding to the overall brightness of the sky
lines was adjusted by fitting the brightest line (the second from the
left) in each spectrum.  The brightest terrestrial line, on the night
when the sky line contamination was the worst (1996 February 24), is
slightly brighter than the M~I cloud (1a--4a) emission.  On all other
nights, the brightest line is $\lesssim$ $\onethird$ of the M~I cloud
intensity.  The analysis of the ``on--off'' pairs was done with and
without this night sky line subtraction to insure that the process was
not introducing spurious emission features.  The results for the
``on--off'' spectra were very similar in the two cases, as is expected
since the sky line intensities change little on hour time scales.

For directions 1a, 4a, 6a, and 12a we also obtained spectra in the
[S~II] $\lambda$6716 line.  On 1996 April 10, two 900 second exposures
were obtained for each of these directions.  

\section{Results}
\label{sec4-results}

\subsection{The H$\alpha$ ``on--off'' Spectra}
\subsubsection{M Complex}
The spectrum obtained by averaging the 5 ``on--off'' pairs measured
for the 1a--4a direction is shown in Figure~\ref{fig4-hvc_1a4a}.  An
emission line is clearly detected at a velocity with respect to the
local standard of rest (LSR) v$_{lsr}$ = -106 $\pm$ 1 km~s$^{-1}$\@.
For v$_{lsr}$ $\leq$ -170 km~s$^{-1}$ the spectrum rises due to our
current inability to accurately correct for the roll-off caused by
vignetting.  For v$_{lsr}$ $\geq$ -50 km~s$^{-1}$ the spectrum
decreases sharply due to differences in the intermediate velocity gas
intensity and the geocoronal intensity between the ``on'' and the
``off'' directions (i.e., the H$\alpha$ emission associated with the
lower velocity gas is brighter in the ``off'' direction).  A Gaussian
fit to the emission line, convolved with the instrumental response
function, is shown as the solid curve in Figure~\ref{fig4-hvc_1a4a}.
The extent of the curve shows the velocity range included in the fit.
The area under this curve leads to an intensity for this emission line
of 0.078 $\pm$ 0.010 R and the full width at half maximum (FWHM) is 27
$\pm$ 4 km~s$^{-1}$.  It must be emphasized that this is the
difference between the emission seen in the 1a direction and the 4a
direction and that the emission spectrum for the 1a direction alone
would be quite different.  The intermediate velocity gas (at least the
neutral portion) is a fairly uniform sheet in this part of the sky,
and so subtracting the 4a direction allows us to extract just the
emission due to the the M~I cloud.  The arrow in the
Figure~\ref{fig4-hvc_1a4a} shows the location of the 21 cm emission
line from neutral hydrogen derived from the Leiden/Dwingeloo survey.
Table~\ref{tab4-obs_results} summarizes the results from all of the
``on--off'' pair analysis.

In order to get a handle on the measurement uncertainties, we fitted
each of the 5 ``on--off'' pairs for the 1a--4a direction separately
and then compared the scatter obtained in the mean, width, and area
parameters of these fits to the uncertainties reported by the fitting
program for these parameters.  The values are in good agreement,
implying that the uncertainties calculated by the fitting program can
be used to estimate the uncertainties involved in measuring the
parameters of the emission lines.  Table~\ref{tab4-obs_results} lists
the calculated uncertainty for each fitted parameter in this and the
other HVC H$\alpha$ observation directions.  It must be emphasized
that these parameters are those of the ``on--off'' spectrum and that
there are potentially significant biases caused by the fact that the
``off'' spectrum contains some unknown quantity of interstellar
H$\alpha$ emission.

The velocities determined for each of the five separate ``on--off''
1a--4a pairs have an average value of -106.1 km s$^{-1}$ with
$\sigma_{v}$ = 1.0 km s$^{-1}$.  This small scatter is a strong
confirmation that the emission line is of Galactic origin, since the
earth's motion along the line of sight changed by approximately 20
km~s$^{-1}$ during the time spanned by the five 1a--4a observations.

Figure~\ref{fig4-hvc_6a12a} shows the 6a--12a difference spectrum.
Although the base line is less well-behaved in this case, an emission
line is again clearly seen at the same velocity as the 21 cm
component.  The intensity of this line is 0.18 $\pm$ 0.02 R, over
twice that found in the 1a direction.  This is interesting since the
H~I column density in the 6a direction is at most $1/6$ (and possibly
0.04) times that in the 1a direction.  Figure~\ref{fig4-map_m} shows
that the 1a direction is located near the peak of the 21 cm
enhancement while the 6a direction is at the edge of the M~I cloud.
The implications of this observation are discussed in more depth in
Section~\ref{sec4-disc}.  A more direct comparison of the H$\alpha$
intensity from directions 1a and 6a could be made by subtracting the
same background direction.  This is impossible in this instance
because there is no ``off'' direction that was observed on both the 6a
night and one of the 1a nights, and one cannot subtract an ``off''
direction spectrum taken on a different night because the sky spectrum
changes and the ``on--off'' technique breaks down.  However, there is
no reason to expect large differences in the high velocity H$\alpha$
between the 4a and the 12a directions since neither has detectable 21
cm emission at the velocity of the M~I cloud.  The available evidence
suggests that they are quite similar, making it unlikely that the
greater intensity seen in the 6a direction is due to the different
``off'' directions that were used.

Figure~\ref{fig4-hvc_2a3a}a shows the 2a--3a spectrum.  The intensity
represented by this emission line is 0.15 $\pm$ 0.02 R, comparable to
the 6a--12a value.  The baseline drops sharply on the red edge of the
line due to bright intermediate velocity gas in the 3a direction (see
below).  The velocity of the H$\alpha$ emission line is well matched
to the 21 cm emission line velocity.  Since spectra of the 4a
direction were obtained on three of the four observation nights, 4a
can be used as a common background direction in many cases to explore
possible uncertainties that might be associated with using a different
``off'' direction in each difference spectrum.
Figure~\ref{fig4-hvc_2a3a}b shows the 2a--4a spectrum.  The 2a spectrum
is the average of spectra derived from two 900 second exposures taken
on February 24, 1996.  On this night we also observed the 4a direction
with two 900 second exposures (see Table~\ref{tab4-obs_info}).  In
general, spectra taken on the same night are close enough in time so
that the sky spectrum can be subtracted out effectively.  The 2a--4a
emission line has an intensity of 0.20 R.  This is significantly more
emission than the 2a--3a line (0.15 R), and the difference is mainly
due to the difference in the fitted widths; the 2a--4a peak height is
only slightly greater than the 2a--3a, and the blue side of the two
lines look quite similar.  This increase in width is a result of there
being less H$\alpha$ emission from intermediate velocity gas in the 4a
direction than in the 3a direction.  This comparison demonstrates that
the results obtained for the emission lines in ``on--off'' pair
spectra do depend somewhat on the ``off'' direction due to the fact
that the ``off'' directions also contain H$\alpha$ emission, but at
generally lower velocities than the HVC detections.

Figure~\ref{fig4-hvc_5a9a}a shows the 5a--9a spectrum.  This spectrum
is fairly noisy, and the baseline is not very well behaved.
Nevertheless, there appears to be an emission line at the same
velocity as the HVC 21 cm emission.  The result of subtracting a 4a
spectrum taken on the same night is shown in
Figure~\ref{fig4-hvc_5a9a}b.  The emission line seen in the 5a--4a
spectrum is somewhat fainter than the one in the 5a--9a spectrum.  The
fit shown in Figure~\ref{fig4-hvc_5a9a}a gives an intensity of 0.09
$\pm$ 0.02 R, while the fit in Figure~\ref{fig4-hvc_5a9a}b corresponds
to an intensity of 0.06 $\pm$ 0.02 R.

Figure~\ref{fig4-hvc_8a4a} shows the the 8a--4a spectrum.  This
emission line has an intensity of 0.1 R, comparable to the 1a--4a
direction.  In this case there appears to be a relatively large (20
km~s$^{-1}$) offset between the H$\alpha$ and 21 cm components.
However, the large difference in widths suggest that there just may be
``extra'' intermediate velocity H$\alpha$ emission in 8a relative to
4a, making the H$\alpha$ line in the difference spectrum appear wider.

The 7a--11a spectrum is displayed in Figure~\ref{fig4-hvc_7a11a}a.
Because 7a is just off the edge of the main H~I cloud and 11a is far
($\simeq$ 16 degrees) from M~I, the non-detection of HVC H$\alpha$ in
this spectrum is evidence against a large halo of ionized gas
surrounding the neutral M~I cloud.  The 7a--4a spectrum
(Fig.~\ref{fig4-hvc_7a11a}b) also shows no high velocity emission
component; if anything, there is a dip, possibly due to the presence
of a small amount of high velocity gas in the 4a direction.

A few of the H$\alpha$ observations were devoted to a search for
ionized gas in the M~II cloud.  Figure~\ref{fig4-hvc_2b3b}a shows the
2b--3b H$\alpha$ spectrum, where 2b is near the peak of the M~II cloud
21 cm enhancement (see Fig.~\ref{fig4-map_m}).  The H$\alpha$ spectrum
shows a clear emission component that is shifted toward less negative
velocities than the M~I cloud components.  This lower velocity is also
apparent in the H~I data.  Figure~\ref{fig4-hvc_2b3b}b shows the result
of subtracting the spectrum of the ``standard'' background direction,
4a, from the 2b spectrum.  In this case, the emission was fitted with
two Gaussian components, since the component has an obvious
non-Gaussian flat-top shape.  The component with a more negative
velocity in this fit is centered at -78 km~s$^{-1}$ and has an
intensity of 0.16 R.  This velocity is not too far from the 21 cm
velocity of -86 km~s$^{-1}$ found for the 2b direction.  The red
component at -50 km~s$^{-1}$ is more in the velocity range normally
considered ``intermediate velocity'' gas.  Once again, because of the
difficulties in isolating the emission components it is not clear
whether there is a real difference between the H$\alpha$ and 21 cm
velocities.  Also, it is interesting to note that our results are in
good agreement with the lower signal to noise M\"{u}nch and Pitz
\markcite{Munch89}(1989) detection of H$\alpha$ emission in a sightline
(l = 185.0\arcdeg, b = 65.0\arcdeg) very near 2b.  They measured a
component at v$_{lsr}$ = -80 km~s$^{-1}$ with I$_{\alpha}$ = 0.15~R\@.
The 1b--4a H$\alpha$ spectrum is shown in Figure~\ref{fig4-hvc_1b4a}.
There is a clear emission component but it is at relatively low
velocity (-61 km~s$^{-1}$), consistent with the lower negative
velocity of the 2b direction.

\subsubsection{A Complex}
Figure~\ref{fig4-hvc_a3}a shows the H$\alpha$ spectrum for the A III
cloud observation ($l$ = 148.5\arcdeg, $b$ = 34.5\arcdeg) resulting
from the average of four 900 second exposures ``on'' the direction
minus the average of four ``off'' observations.  Notice that the
spectrometer was ``tuned'' to cover a higher negative velocity
interval than for the M complex observations.  This is accomplished by
changing the optical path length in the Fabry-Perot etalon gaps by
changing the gas pressure in the etalon chambers.  The spectrum shows
a clear detection of an emission line at a velocity of v$_{lsr}$ =
-167 $\pm$ 1 km~s$^{-1}$.  The arrow in the figure shows the velocity
of the component detected for the 21 cm line in the Leiden/Dwingeloo
survey data, which appears to have a significant offset from the
H$\alpha$ velocity.  The H$\alpha$ intensity derived from the fit
shown in Figure~\ref{fig4-hvc_a3}a is I$_{\alpha}$ = 0.08 $\pm$ 0.01
R\@.

Figure~\ref{fig4-hvc_a3}b shows the H$\alpha$ spectrum measured for
the A IV cloud direction ($l$ = 153.6\arcdeg, $b$ = 38.2\arcdeg), the
average of four 900 second ``on'' exposures minus the average of three
900 second ``off'' exposures.  The emission feature detected in this
case has an intensity of I$_{\alpha}$ = 0.09 $\pm$ 0.01 R and a
velocity of v$_{lsr}$ = -178 $\pm$ 1 km~s$^{-1}$.  The arrow in the
figure shows the velocity of the 21 cm emission, which does match the
H$\alpha$ component velocity.

\subsubsection{C Complex}
Figure~\ref{fig4-hvc_c} shows the H$\alpha$ difference spectrum for
the C complex direction ($l$ = 84.3\arcdeg, $b$ = 43.7\arcdeg).  While
the baseline is not well determined, an emission component is clearly
detected at a velocity matching that of the 21 cm component.  The
H$\alpha$ intensity of this component is I$_{\alpha}$ = 0.13 $\pm$ 0.03
R, and the velocity is v$_{lsr}$ = -111 $\pm$ 2 km~s$^{-1}$.
Songaila, Bryant, \& Cowie \markcite{Songaila89}(1989) observed this
same look direction (their Position 1, see their Fig. 1a).  They
report a detection with an intensity of 0.03 R and an estimated
uncertainty of 50\%.  Our higher signal to noise observation is
inconsistent with this result.

\subsection{The [S~II] $\lambda$6716 ``on--off'' Spectra}
Figure~\ref{fig4-sii_1a4a}a shows the [S~II] $\lambda$6716 spectrum of
the 1a--4a direction in the M~I cloud.  The spectrum strongly suggests the
presence of [S~II] emission at a velocity very close to the velocity
at which H$\alpha$ emission was detected in this direction (indicated
by an arrow in the figure).  The best fit intensity is 0.05 $\pm$ 0.01
R, and the width is 23 $\pm$ 8 km~s$^{-1}$ (the uncertainties quoted
here are the formal errors calculated in the fitting process).  The
mean velocity is v$_{lsr}$ = -110 $\pm$ 3 km~s$^{-1}$.  The spectrum
also suggests another smaller peak near -60 km~s$^{-1}$ (not fitted).
Figure~\ref{fig4-sii_1a4a}b shows the [S~II] $\lambda$6716 spectrum of
the 6a--12a direction.  This spectrum does not show a clear [S~II]
detection; however, there is a marginal enhancement at -96
km~s$^{-1}$.  The Gaussian fit to this enhancement corresponds to an
intensity of 0.02 $\pm$ 0.01 R and a width of 18 km~s$^{-1}$.  We
will treat the fitted intensity as an upper limit to the [S~II]
emission from this direction.  The [S~II] / H$\alpha$ ratios for the
two directions are then:
\begin{equation}
\left(\frac{I_{\rm S~II}}{I_{H\alpha}}\right)_{1a4a} \simeq 0.64 \pm 0.14,
\end{equation}
and
\begin{equation}
\left(\frac{I_{\rm S~II}}{I_{H\alpha}}\right)_{6a12a} \lesssim 0.11 \pm
0.06,
\end{equation}
a difference of a factor of six.  This large difference in the [S~II]
/ H$\alpha$ ratio indicates significant variations in
ionization/excitation conditions within the M~I cloud.

\section{Discussion and Conclusions}
\label{sec4-disc}

We have searched for H$\alpha$ emission from five high velocity clouds
in three cloud complexes.  In every direction where there is
significant high velocity H~I gas seen in 21 cm emission, an
associated H$\alpha$ line was detected.  Also, except for 6a, which is
just off the edge of the M~I cloud, there is no clear evidence of
H$\alpha$ in any of the directions without appreciable H~I emission.
Nevertheless, while there is a close correspondence between high
velocity 21 cm and H$\alpha$ emission on the sky, the intensities of
the 21 cm and H$\alpha$ are not well correlated.
Figure~\ref{fig4-inten_corr} shows the H$\alpha$ intensity in
Rayleighs from Table~\ref{tab4-obs_results} plotted versus the 21 cm
column density determined from the Leiden/Dwingeloo data for the M
complex directions.  No correlation is apparent in the plot.  The
brightness of the 6a direction provides some evidence that the ionized
gas producing the H$\alpha$ emission envelopes the 21 cm emitting
neutral gas. However, the H$\alpha$ ``halo'', if present, is not
large, as is shown, for example, by the nondetection of H$\alpha$ in
the 7a direction.  A complete map of the high velocity H$\alpha$
emission in this region would be very illuminating in this regard.

On the other hand, the velocities of the components found in the
H$\alpha$ spectra do correlate with the velocities found in the 21 cm
data.  Figure~\ref{fig4-vel_corr} shows a plot of the H$\alpha$
component velocities versus the 21 cm component velocities.  A good
correlation between the two exists, indicating that the H$\alpha$
emission is clearly associated with the H~I clouds seen in the 21 cm
line.  The scatter is consistent with our current measurement errors.
The systematic offset (of -10 km s$^{-1}$) of the H$\alpha$ from the
21 cm is tantalizing, but may well also be due to biases introduced by
the ``on--off'' subtraction method (see below).

The widths of the H$\alpha$ components are very uncertain.  This is
due to the fact that in many cases the high velocity emission is
blended with intermediate velocity emission, which the ``on--off''
method subtracts out, but not without introducing biases into the
measured line widths (and therefore also positions) of the high
velocity emission component.  The 1a--4a direction provides the best
measurement we have of a high velocity emission component, and in this
direction we measure the width to be 27 $\pm$ 4 km~s$^{-1}$.  This is
quite similar to the 24 km~s$^{-1}$ line width measured for the 21 cm
emission component in this direction (see
Table~\ref{tab4-obs_results}).  The bias due to the ``off''
subtraction can be seen by comparing the width found for the 2a--3a
spectrum (W = 38 km~s$^{-1}$) to that measured for the 2a--4a spectrum
(W = 49 km s$^{-1}$).  As another example, the 5a--9a emission line
has a width of 46 km~s$^{-1}$, while the 5a--4a emission line is 38
km~s$^{-1}$ wide.  Therefore, the component widths are not determined
well, and the width uncertainties listed in
Table~\ref{tab4-obs_results}, which were calculated from the quality
of the fits and without considering the bias due to the ``off''
subtraction, are underestimated.  The [S~II] line width in the 1a
direction appears to be narrower than the H$\alpha$ line width, but is
highly uncertain.  The formal fit gives 23 $\pm$ 8 km~s$^{-1}$ (FWHM)
for the [S~II] line width compared to 27 $\pm$ 4 km~s$^{-1}$ for the
H$\alpha$ line.

Jura \markcite{Jura79}(1979) has suggested the possibility of
detecting high velocity H$\alpha$ emission that originates in the
plane of the Galaxy and is reflected off of dust contained in HVCs.
This emission would have radial velocities approximately twice that of
the 21 cm emission and hence, this is not the origin of the the
emission line detections reported here.  The velocities expected for
this reflected emission are unfortunately outside the range probed in
this study.  However, a search for it in a future study is a viable
and interesting possibility that could provide unique information on
the optical transparancy of the Galactic disk as well as further
measurements of the dust content of the HVCs.

\begin{figure}[p]
  \centerline{ \epsfysize = 6.0in \epsffile{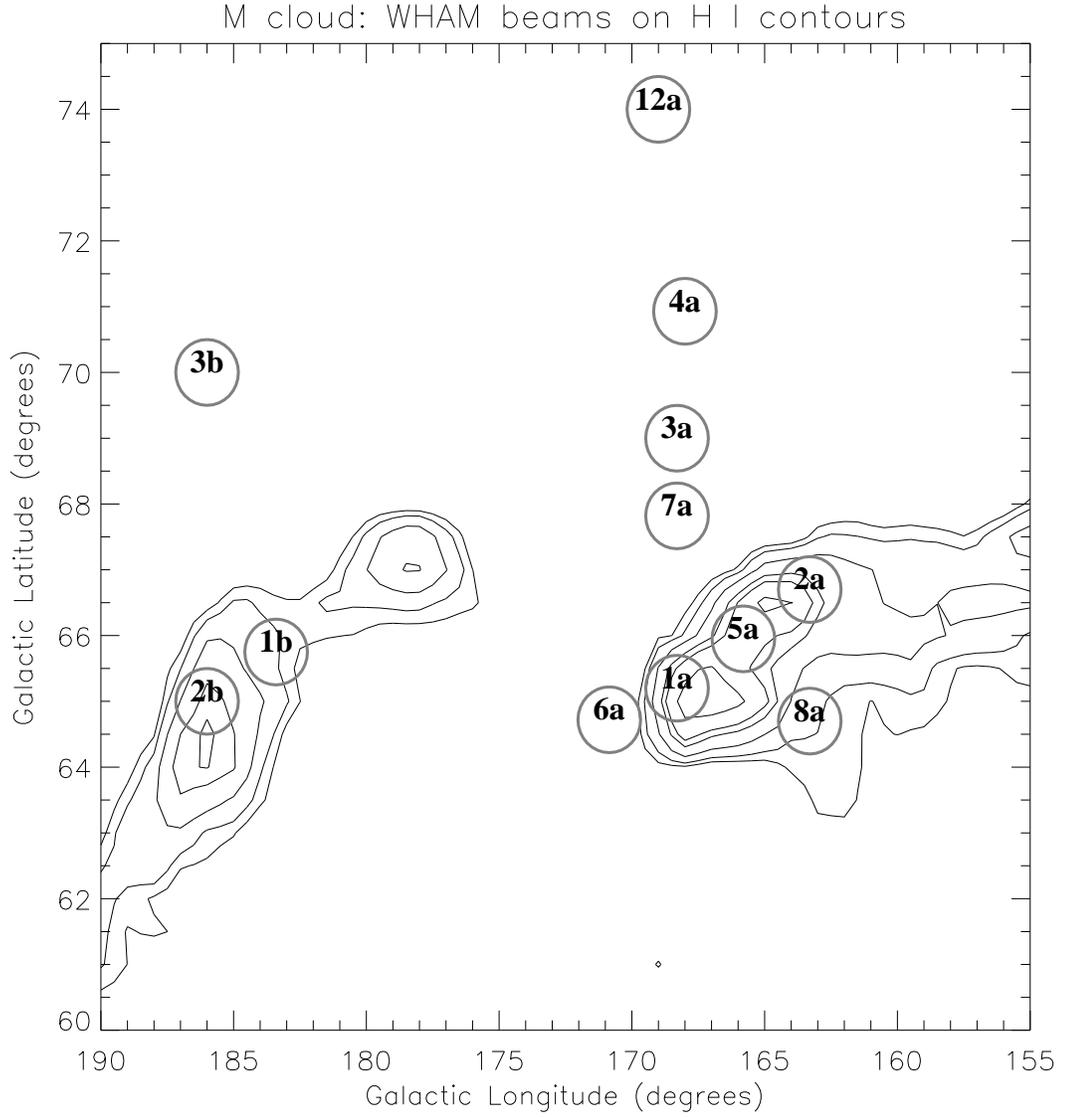} } \caption{WHAM
    observation directions for the M cloud superposed on 21 cm
    contours.  The 21 cm intensity contours correspond to v$_{lsr}$
    $<$ -80 km s$^{-1}$.  The contour levels correspond to N$_{\rm H
    I}$ = 2.0, 3.0, 5.0, 8.0, 10.0, and 15.0 $\times$ 10$^{19}$
    cm$^{-2}$.}
\label{fig4-map_m}
\end{figure}

\begin{figure}[t]
  \centerline{ \epsfysize = 7.0in \epsffile{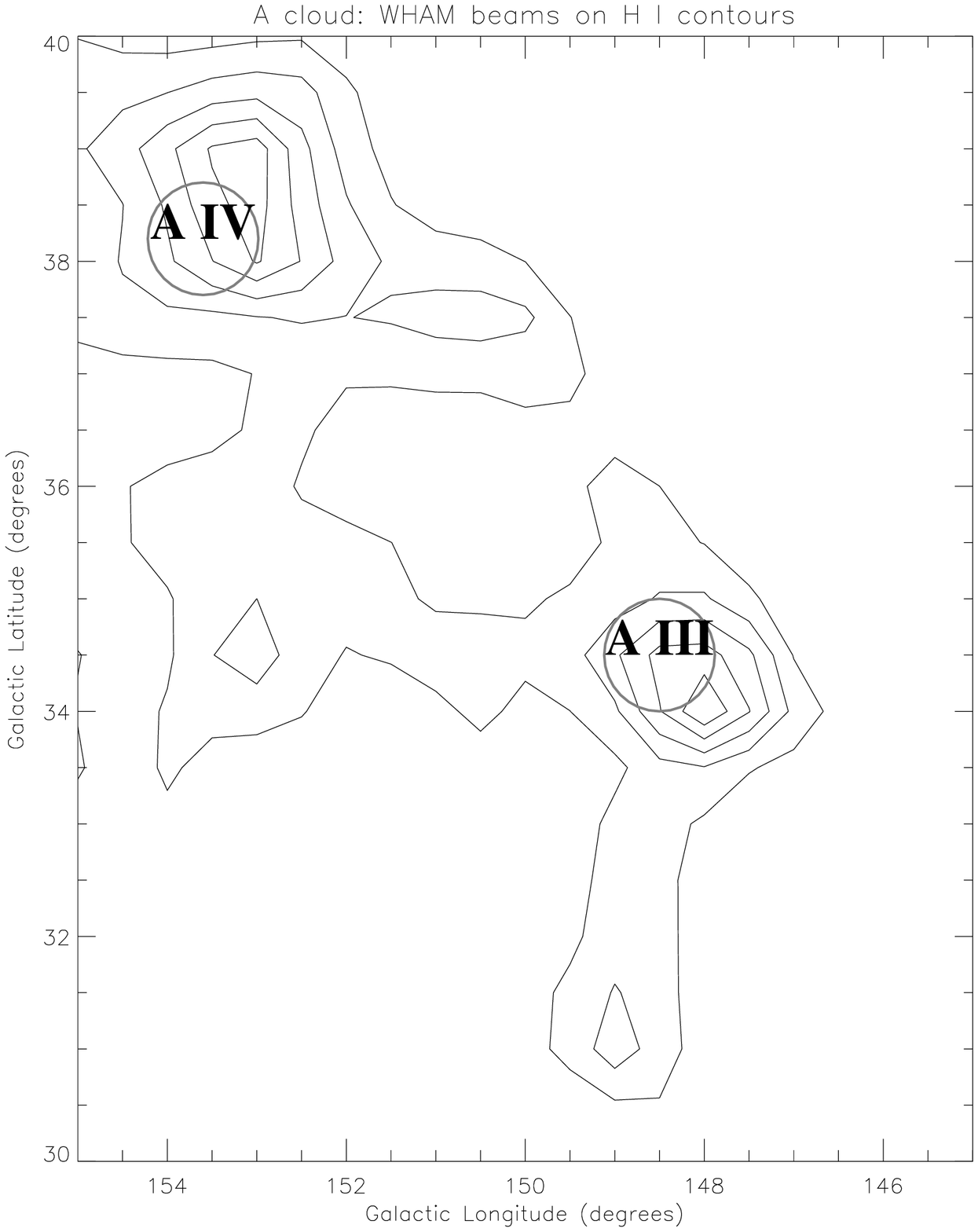} } \caption{WHAM
    observation directions for the A cloud superposed on 21 cm
    contours (for v$_{lsr}$ $<$ -100 km s$^{-1}$).}
\label{fig4-map_a}
\end{figure}

\begin{figure}[t]
  \centerline{ \epsfysize = 3.25in \epsffile{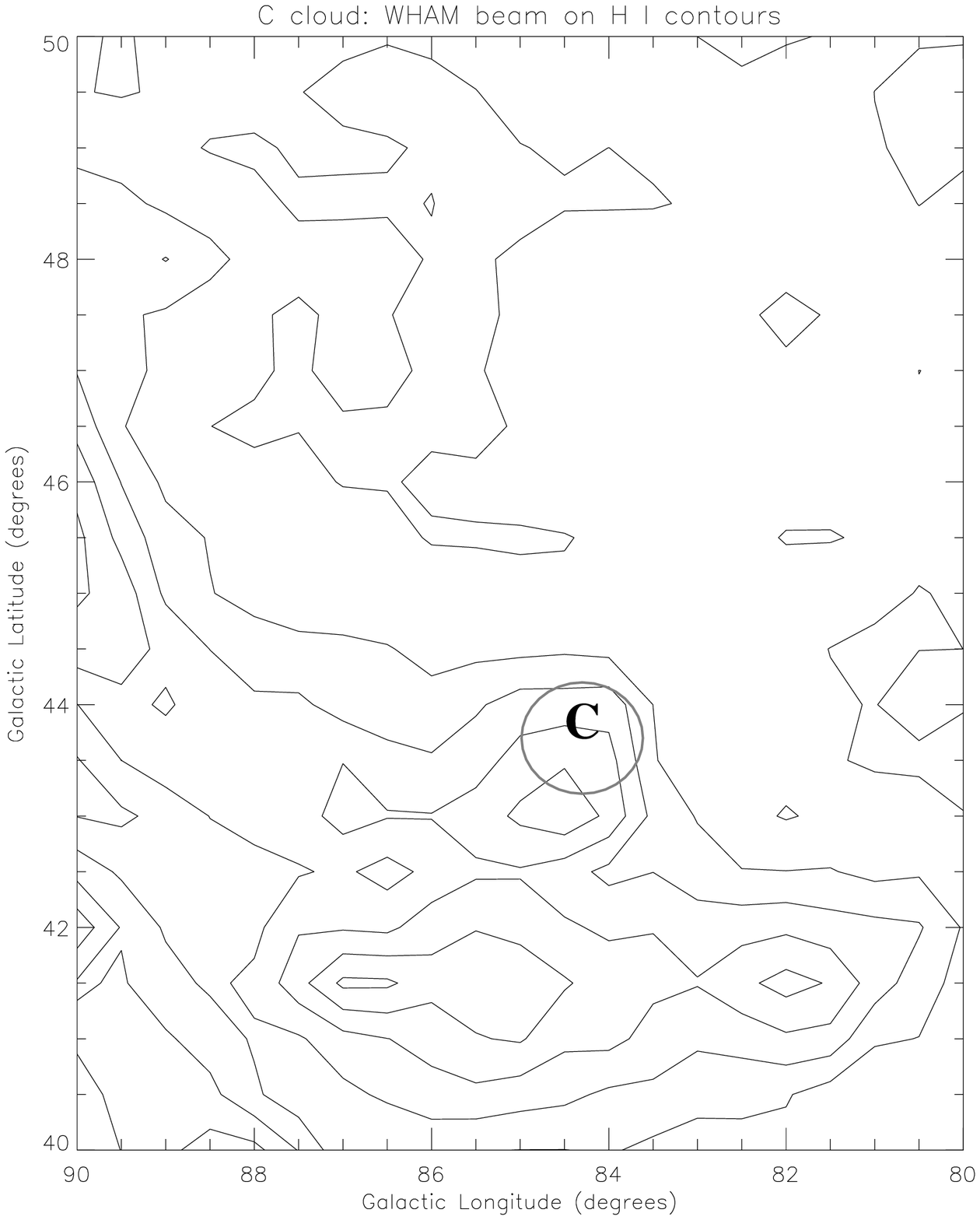} } \caption{WHAM
    observation direction for the C cloud superposed on 21 cm contours
    (for v$_{lsr}$ $<$ -100 km s$^{-1}$).}
\label{fig4-map_c}
\end{figure}

\begin{figure}[p]
  \centerline{
    \epsfysize = 6.5in
    \epsffile{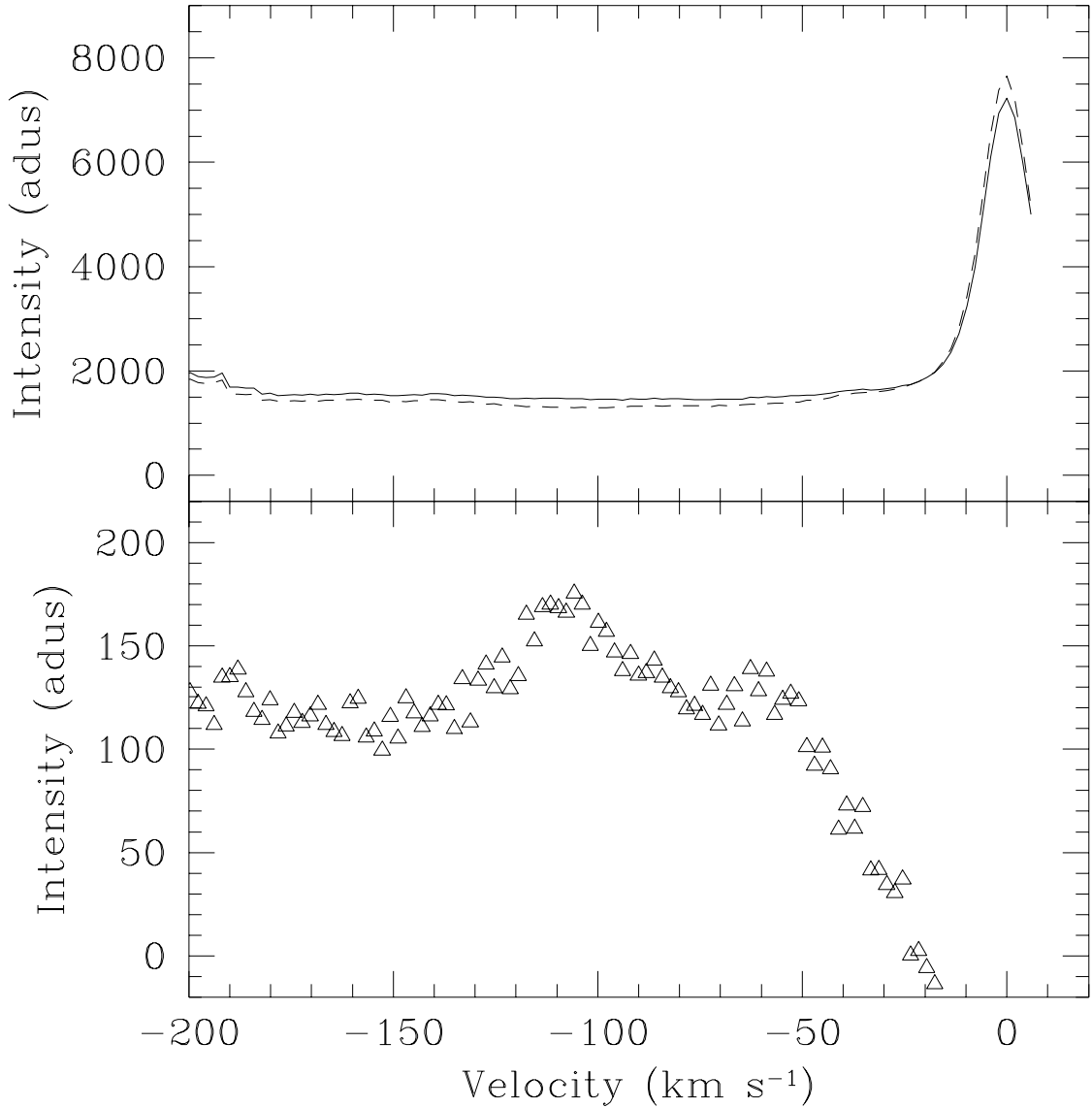}
    }
  \caption{H$\alpha$ from the M~I cloud: the upper panel shows the spectra
    derived from single 900 second exposures ``on'' (the solid line is
    direction 1a) and ``off'' (dashed line is direction 4a) the M~I
    cloud.  The lower panel shows the difference (``on - off'') spectrum.
    Note that the units on the intensity axis are arbitrary 
    (adus = ``arbitrary data units'').}
\label{fig4-hvc_on_off}
\end{figure}

\begin{figure}[t]
  \centerline{
    \epsfysize = 6.5in
    \epsffile{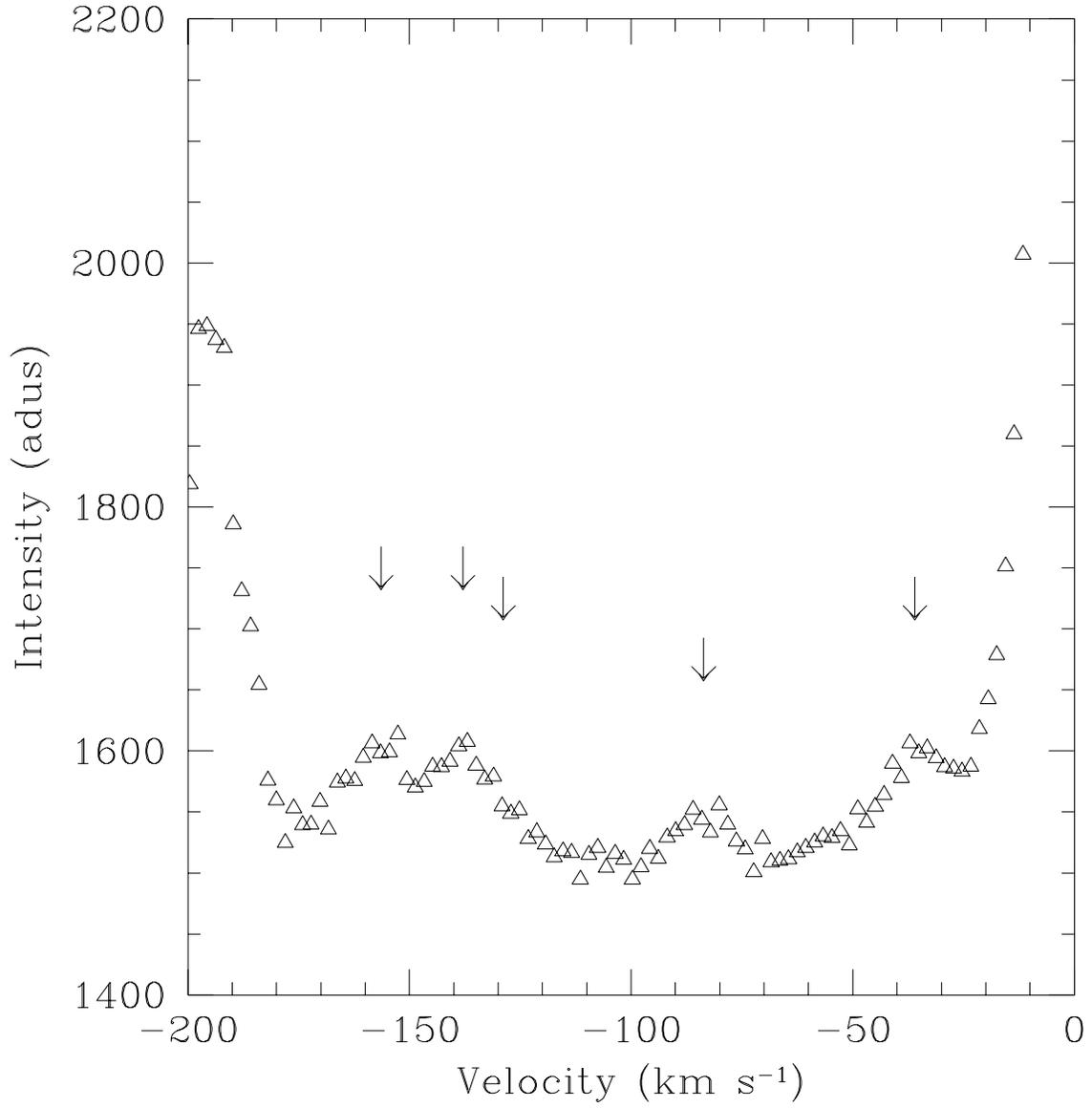}
    }
  \caption{Night sky spectral features that are present with variable 
    strength in all the HVC spectra.}
\label{fig4-oh_spect}
\end{figure}

\begin{figure}[t]
  \centerline{
    \epsfysize = 3.25in
    \epsffile{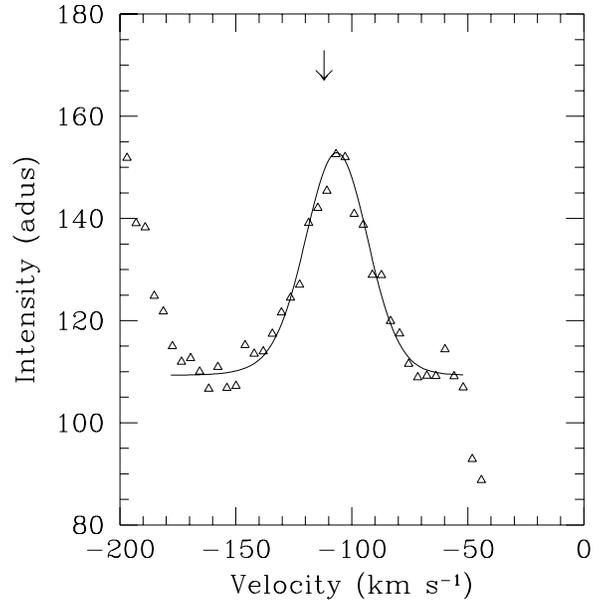}
    }
  \caption{H$\alpha$ from the M~I cloud: 1a - 4a (normalized to 900s 
    exposure time).  The arrow shows the velocity of the 21 cm 
    HVC detection.}
\label{fig4-hvc_1a4a}
\end{figure}

\begin{figure}[b]
  \centerline{
    \epsfysize = 3.25in
    \epsffile{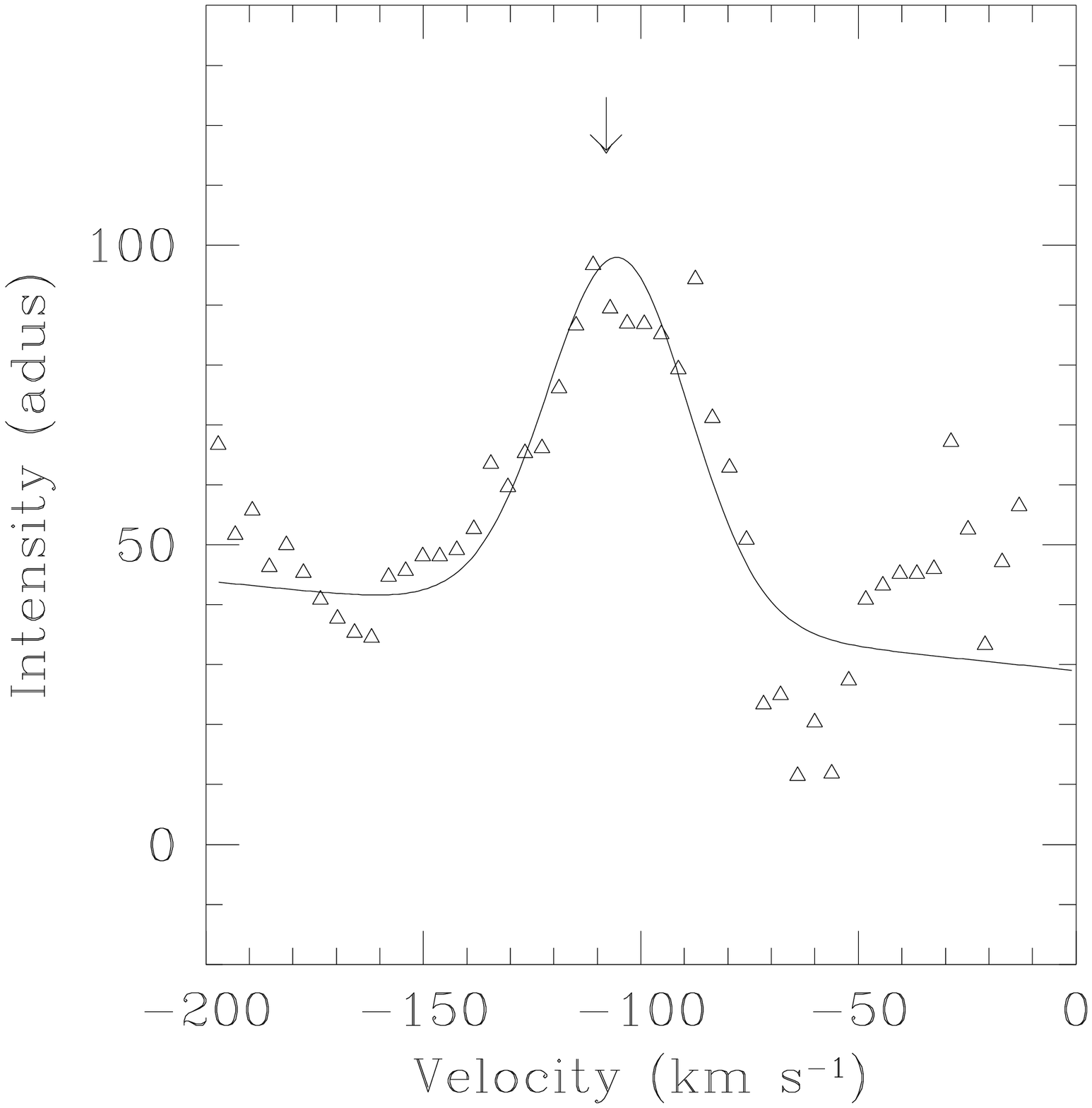}
    }
  \caption{H$\alpha$ from the M~I cloud: 6a - 12a (900s exposure).  The arrow
    shows the velocity of the 21 cm HVC detection.}
\label{fig4-hvc_6a12a}
\end{figure}

\begin{figure}[t]
  \plottwo{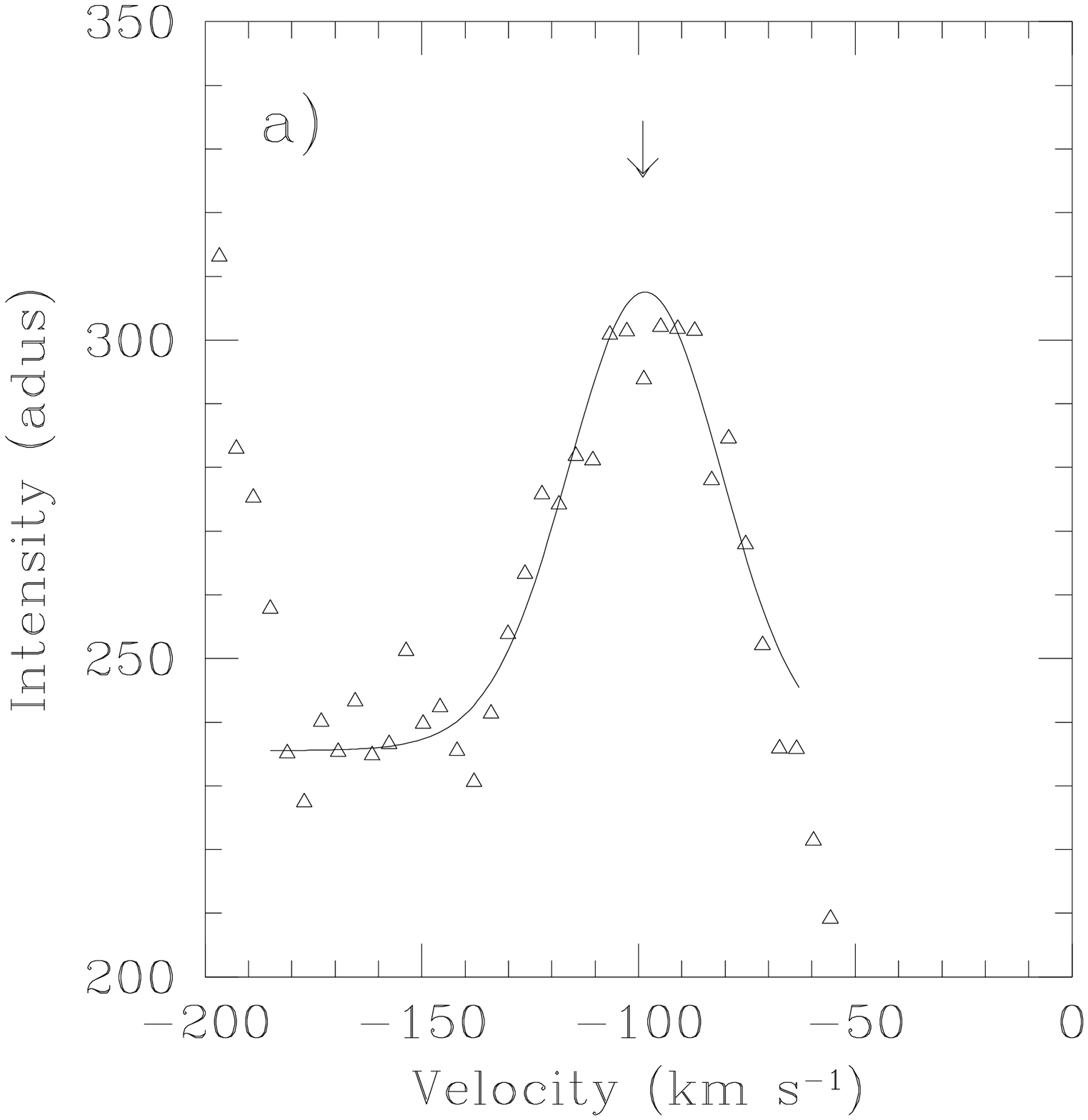}{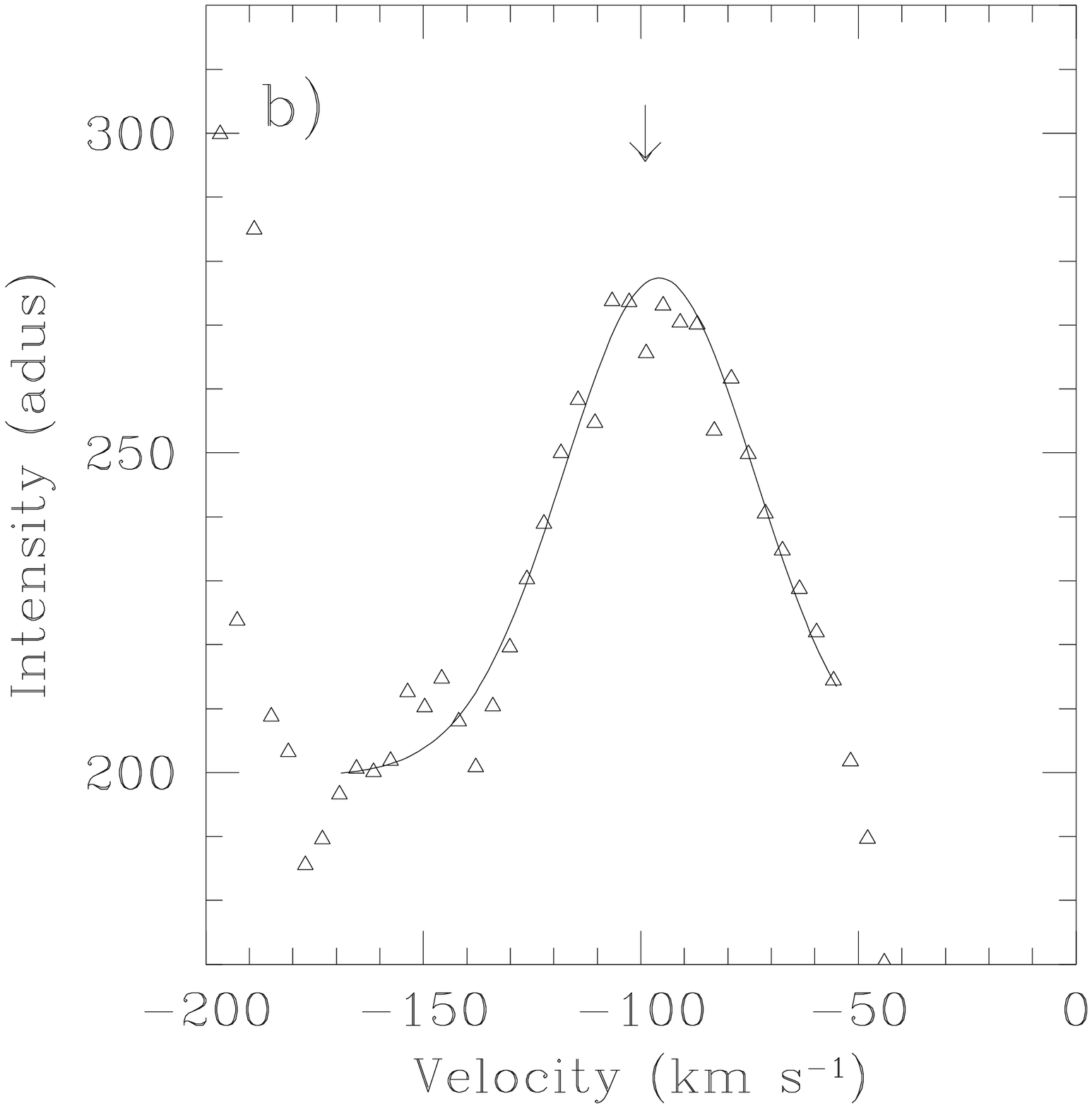}
  \caption{H$\alpha$ from the M~I cloud: a) 2a - 3a.  b) 2a - 4a
  (both normalized to 900s exposure).  The arrows show the velocity of
  the 21 cm HVC detections.}
\label{fig4-hvc_2a3a}
\end{figure}

\begin{figure}[b]
  \plottwo{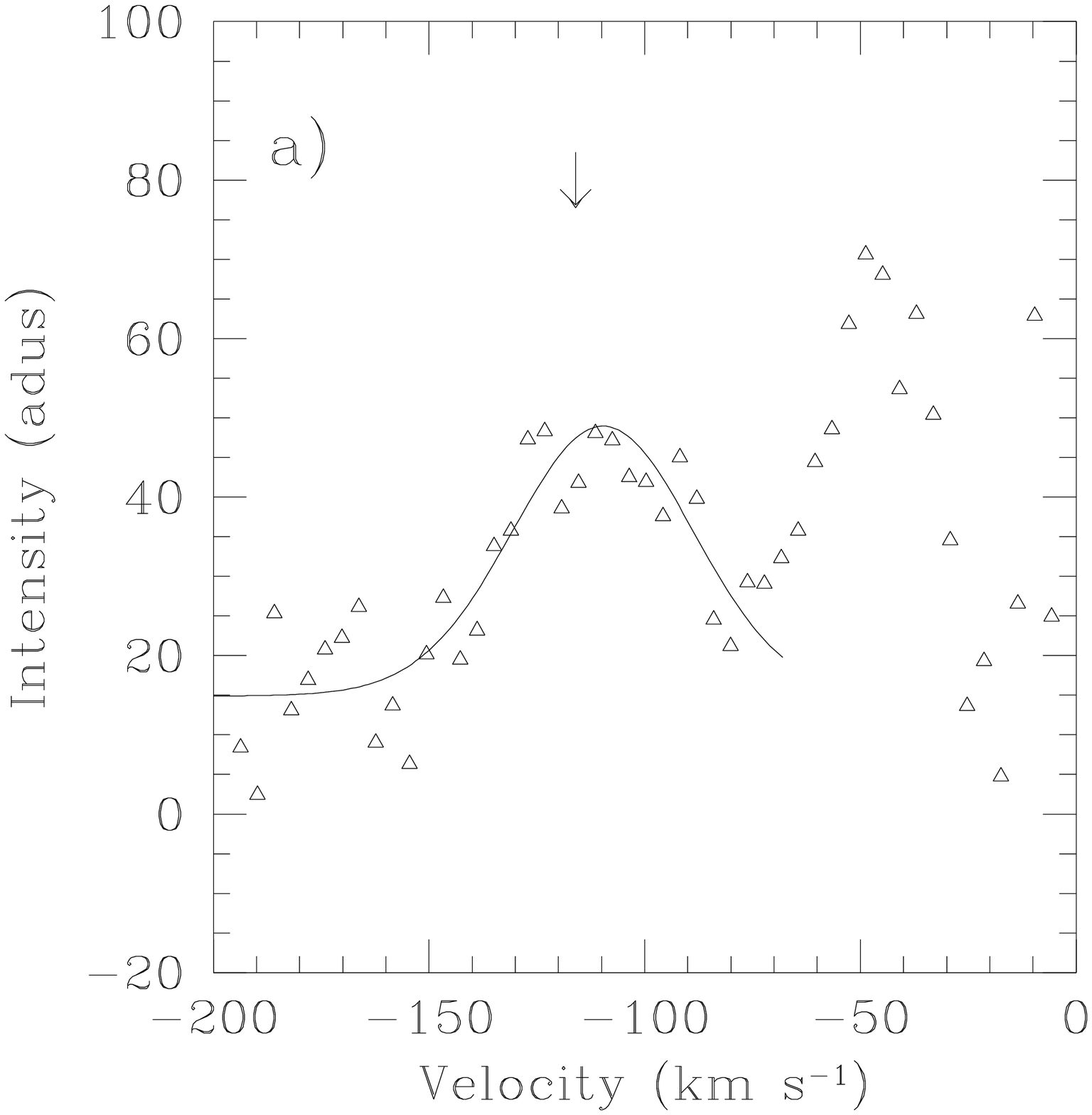}{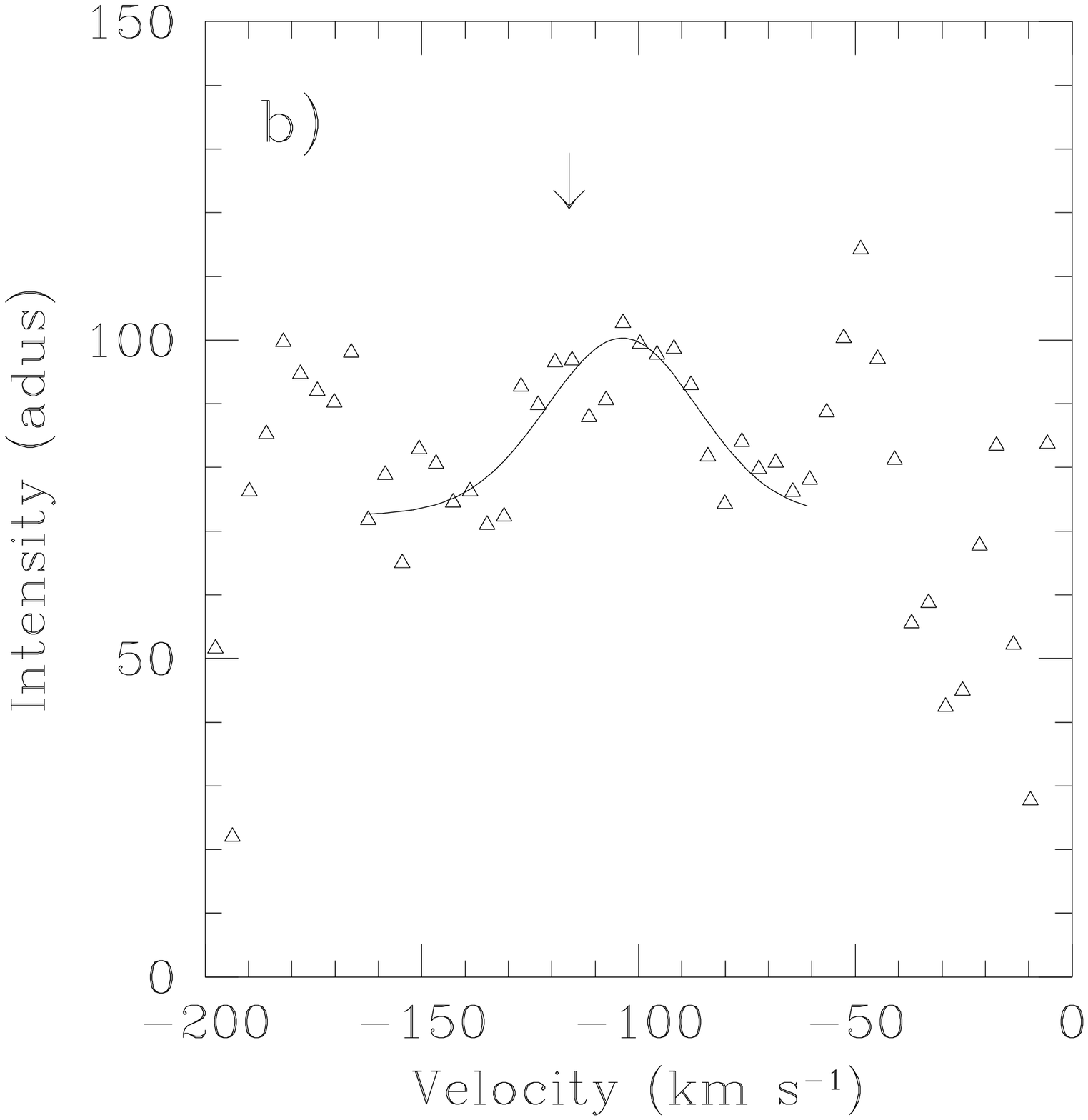}
  \caption{H$\alpha$ from the M~I cloud: a) 5a - 9a.  b) 5a - 4a  (both 
  are 900s exposures).  The arrows show the velocity of the 21 cm HVC 
  detections.}
\label{fig4-hvc_5a9a}
\end{figure}

\begin{figure}[t]
  \centerline{
    \epsfysize = 3.25in
    \epsffile{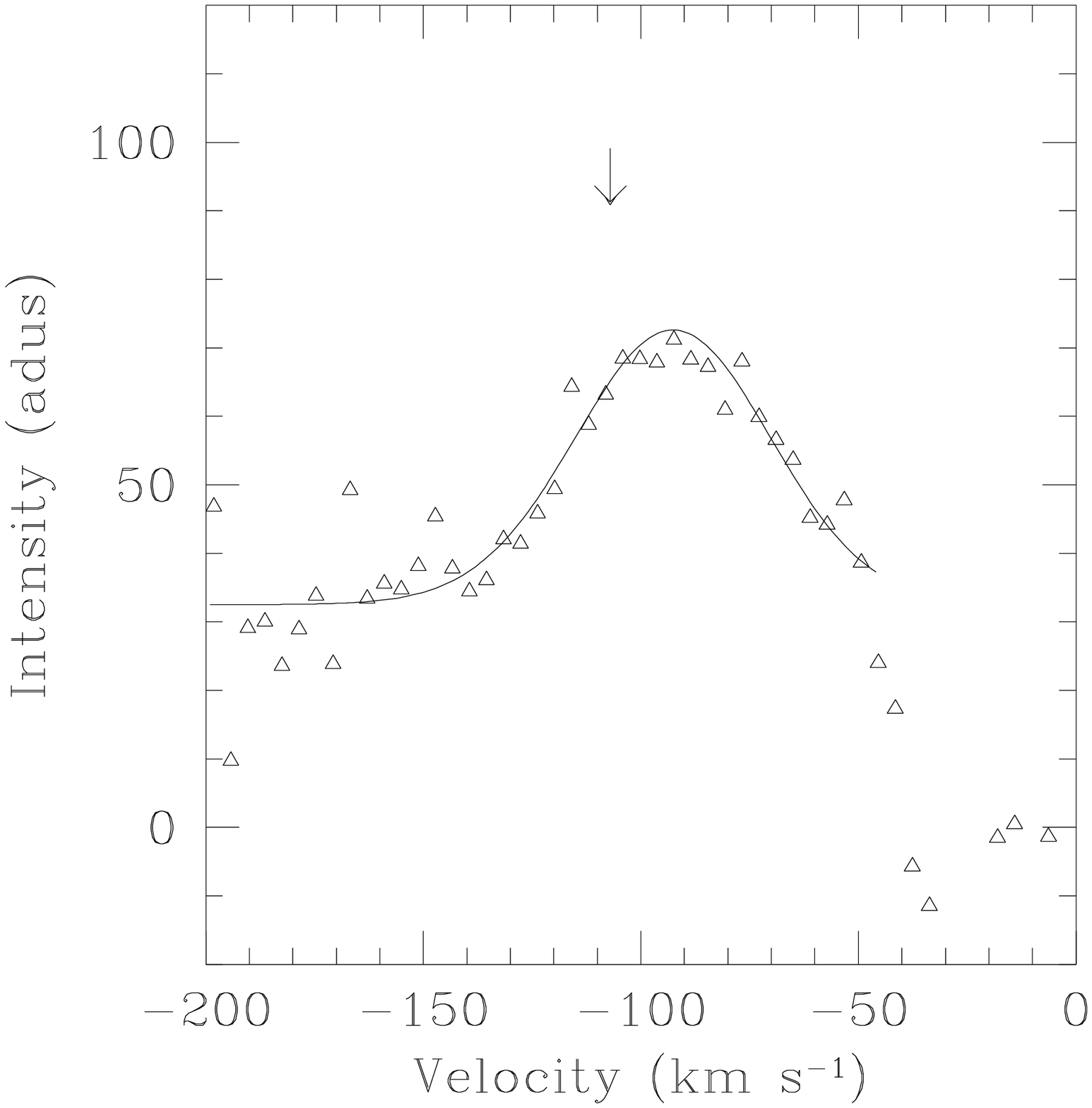}
    }
  \caption{H$\alpha$ from the M~I cloud: 8a - 4a (900s exposure).  The arrow
    show the velocity of the 21 cm HVC detection.}
\label{fig4-hvc_8a4a}
\end{figure}

\begin{figure}[b]
  \plottwo{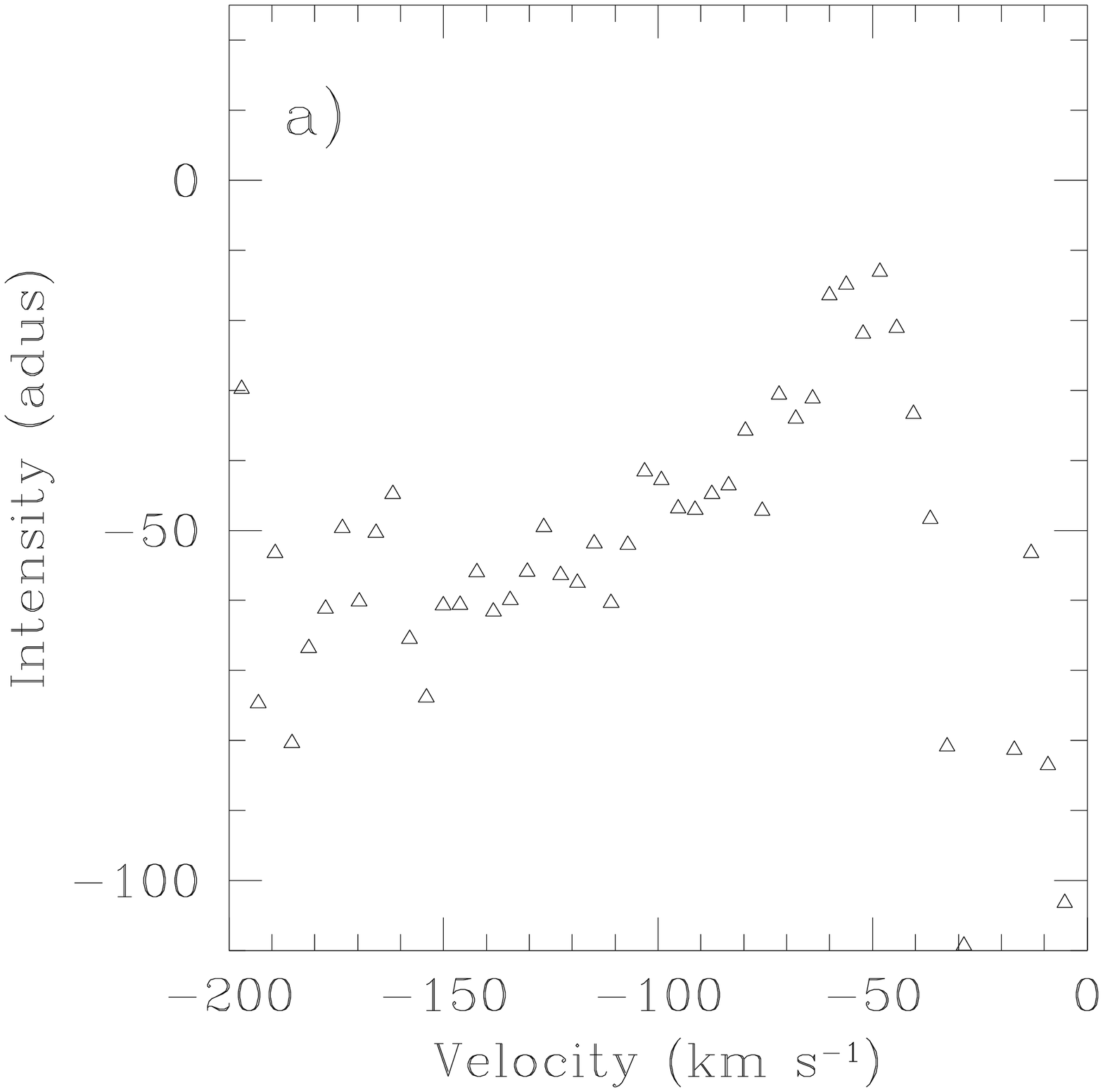}{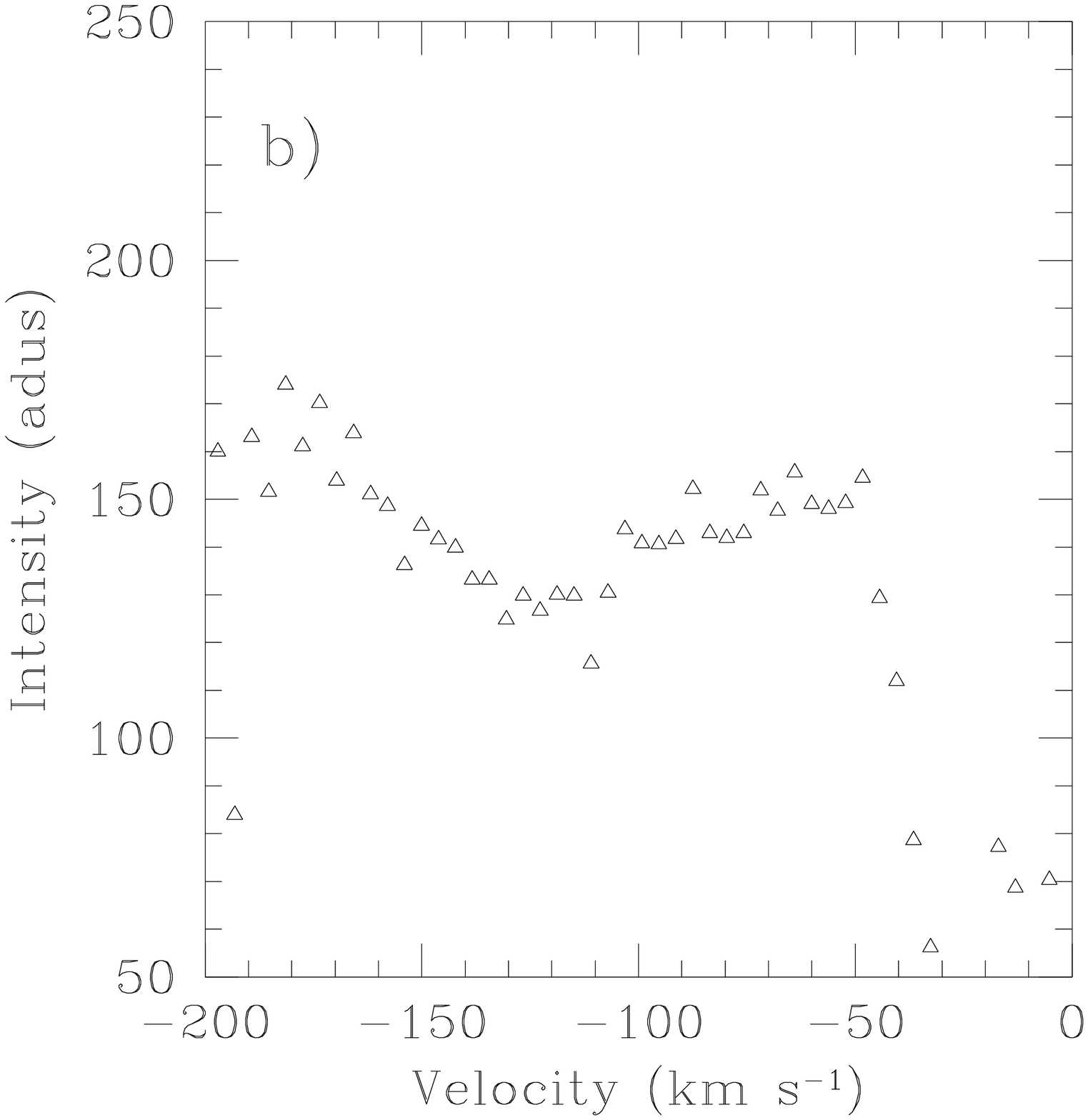}
  \caption{H$\alpha$ from the M~I cloud: a) 7a - 11a.  b) 7a - 4a (both 
  are 900s exposures).}
\label{fig4-hvc_7a11a}
\end{figure}

\begin{figure}[t]
  \plottwo{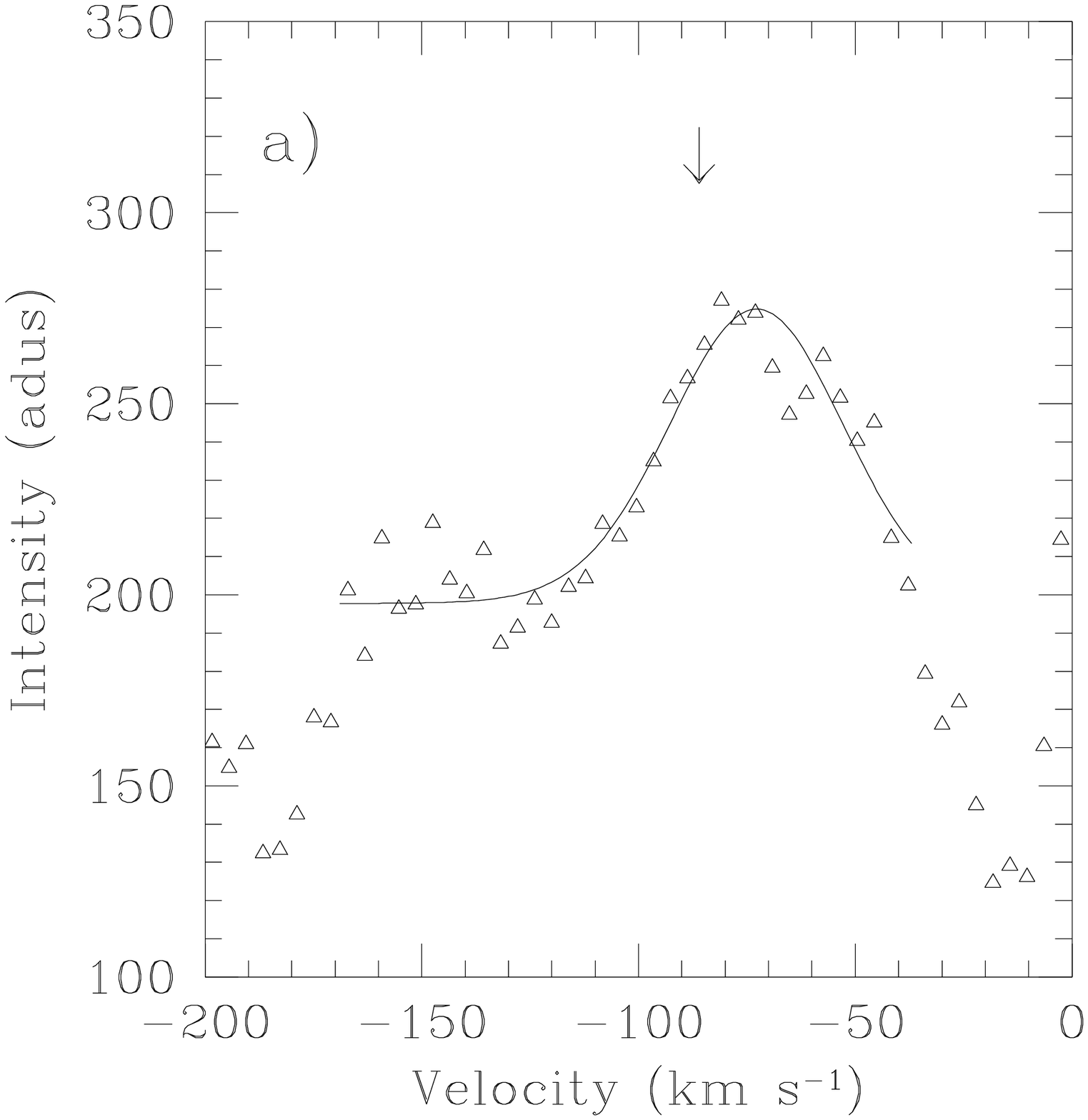}{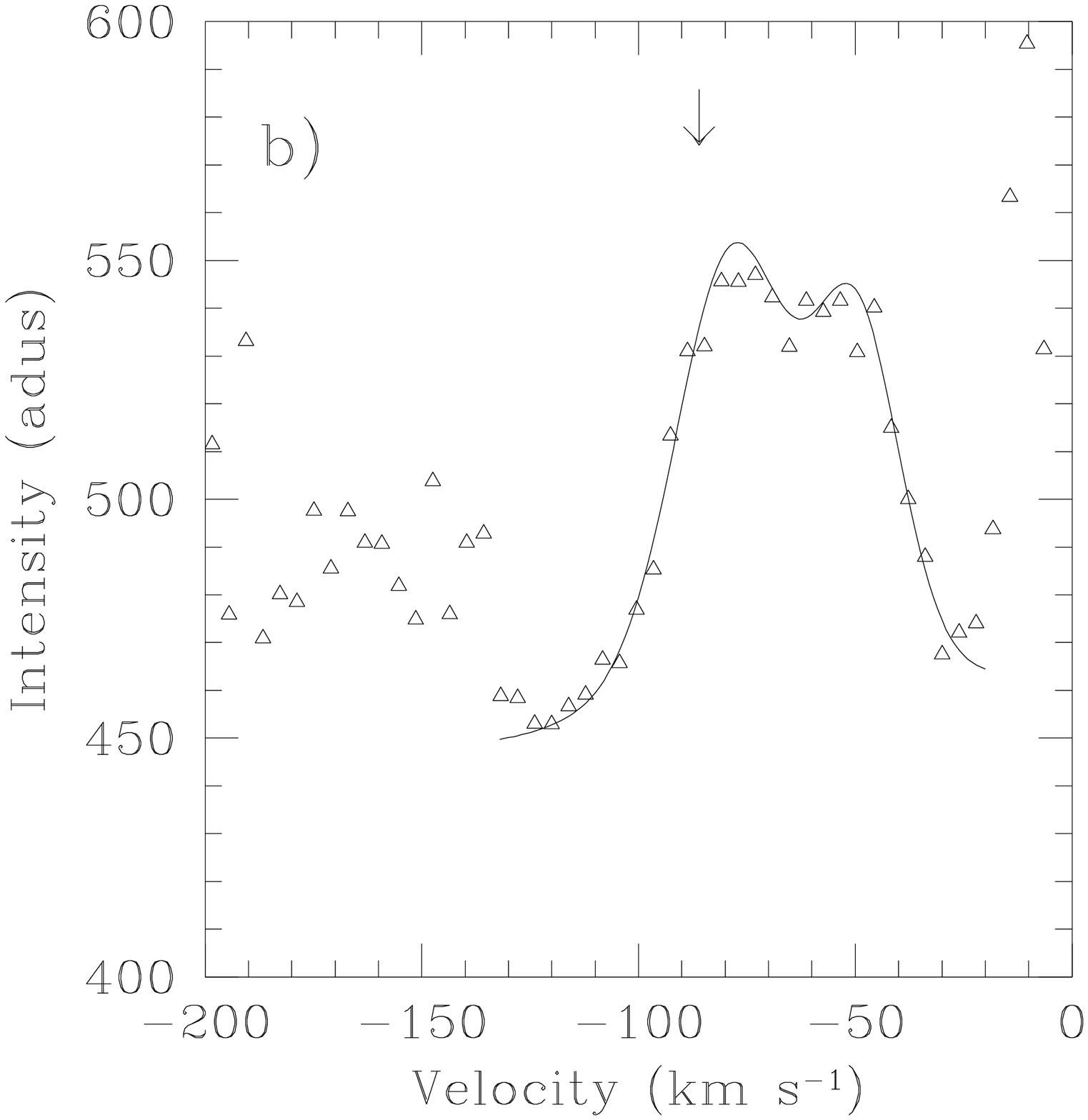}
  \caption{H$\alpha$ from the M~II cloud: a) 2b - 3b.  b) 2b - 4a (both 
  are 900s exposures).  The arrows show the velocity of the 21 cm HVC 
  detections.}
\label{fig4-hvc_2b3b}
\end{figure}

\begin{figure}[t]
  \centerline{
    \epsfysize = 3.25in
    \epsffile{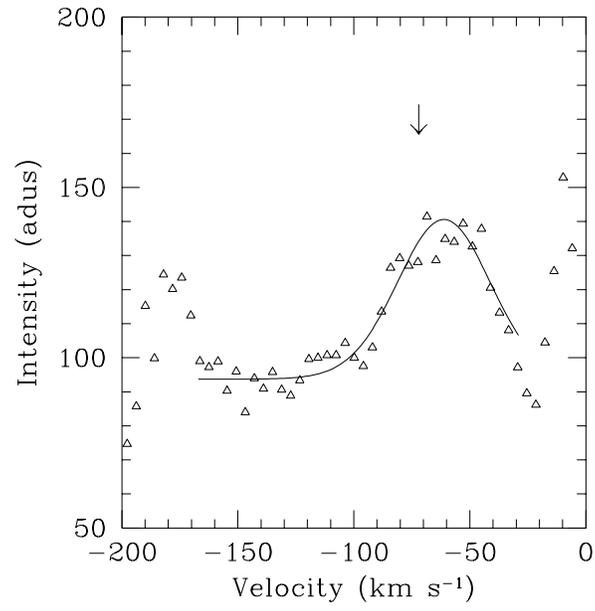}
    }
  \caption{H$\alpha$ from the M~II cloud: 1b - 4a (900s exposure).}
\label{fig4-hvc_1b4a}
\end{figure}

\begin{figure}[t]
  \plottwo{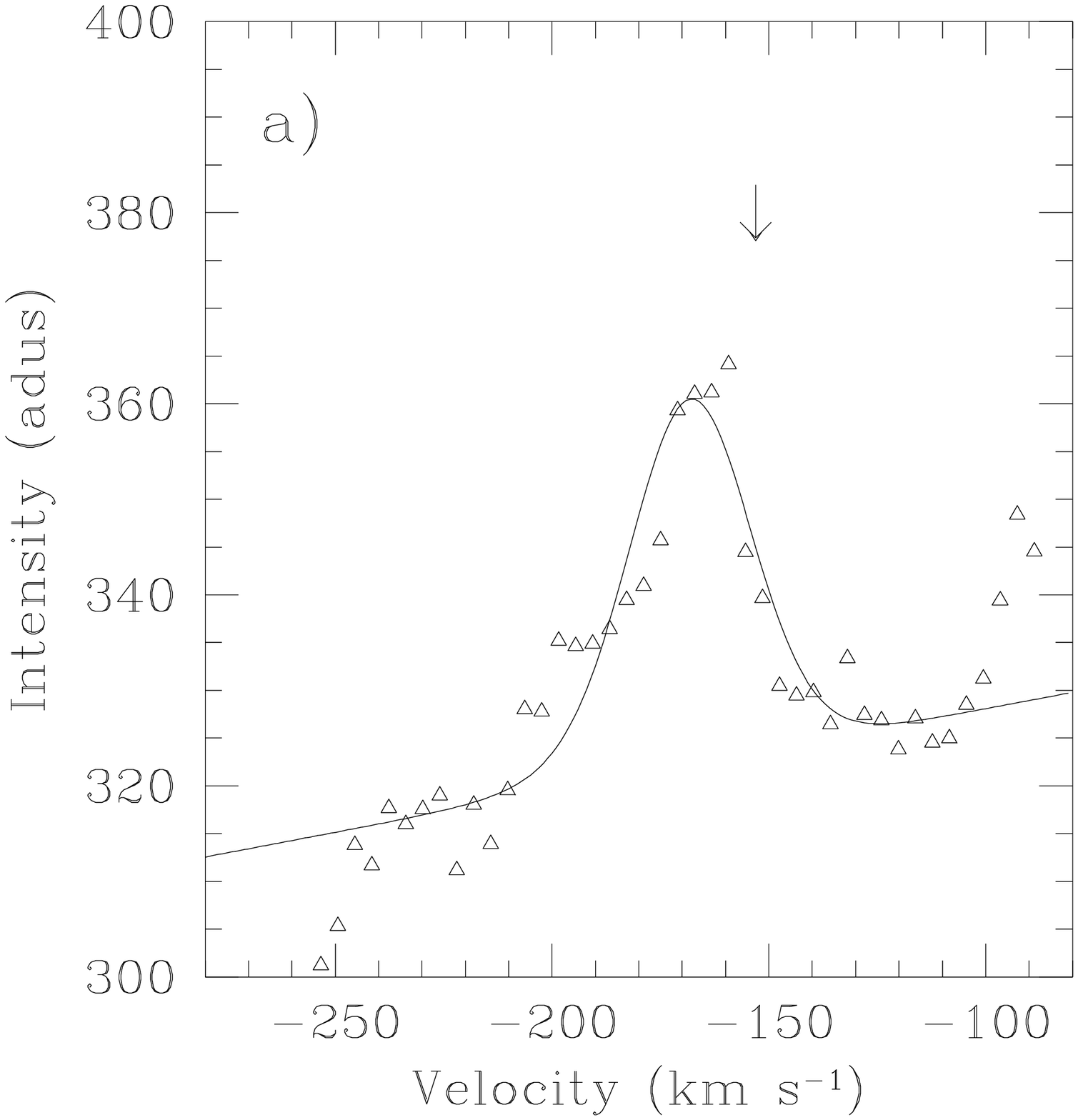}{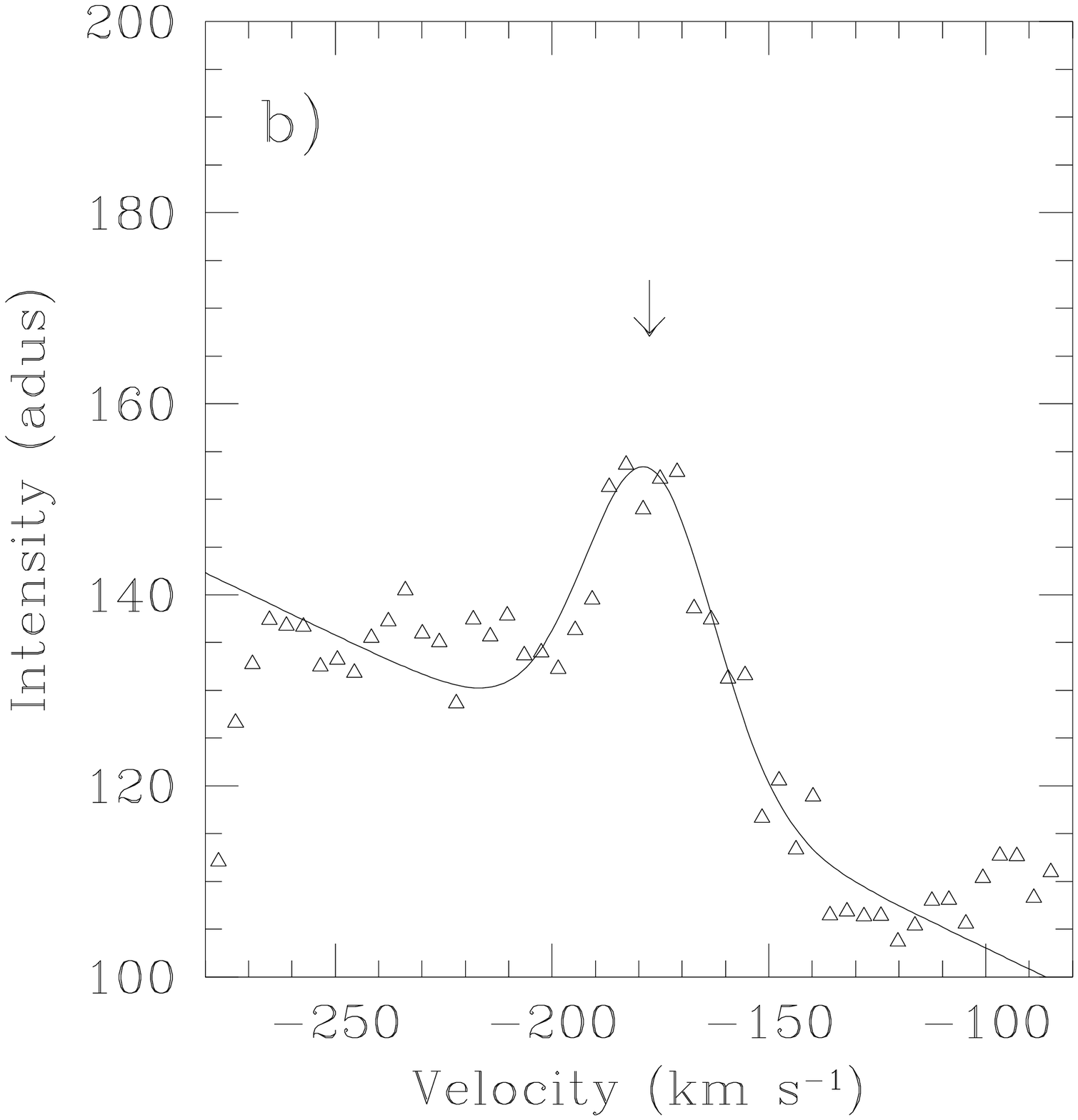}
  \caption{H$\alpha$ from the A Complex.  a) A III cloud.  b) A IV cloud 
   (both normalized to 900s exposure).}
\label{fig4-hvc_a3}
\end{figure}

\begin{figure}[b]
  \centerline{ 
    \epsfysize = 3.25in 
    \epsffile{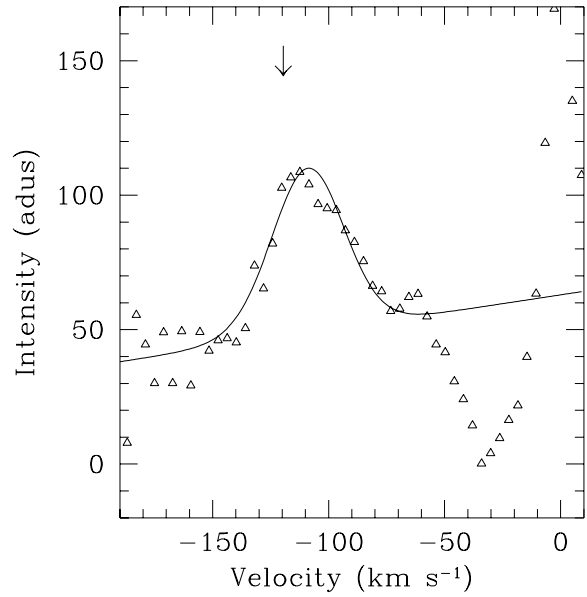} }
  \caption{H$\alpha$ from the C cloud (normalized to 900s exposure).}
\label{fig4-hvc_c}
\end{figure}

\begin{figure}[t]
  \plottwo{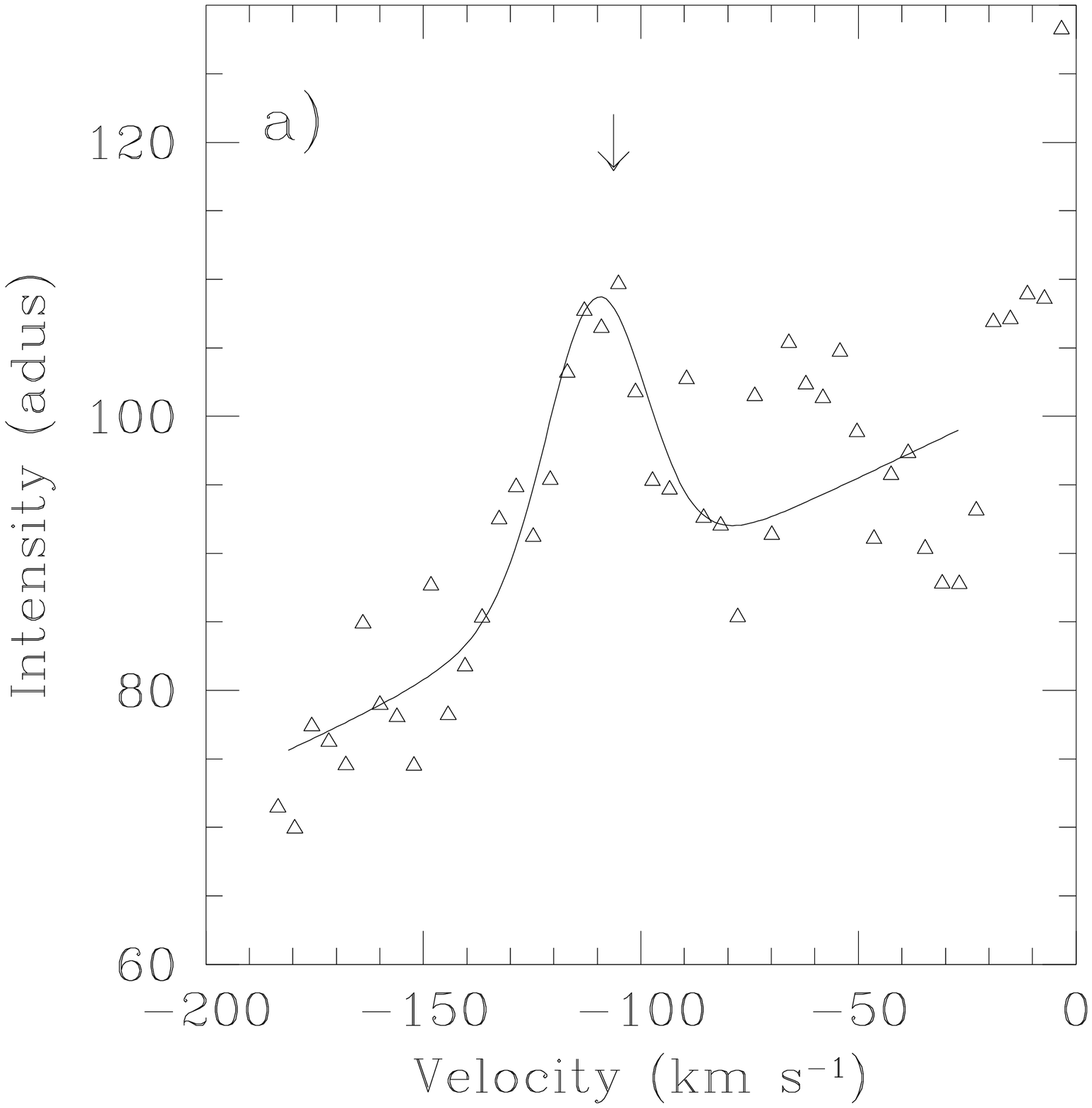}{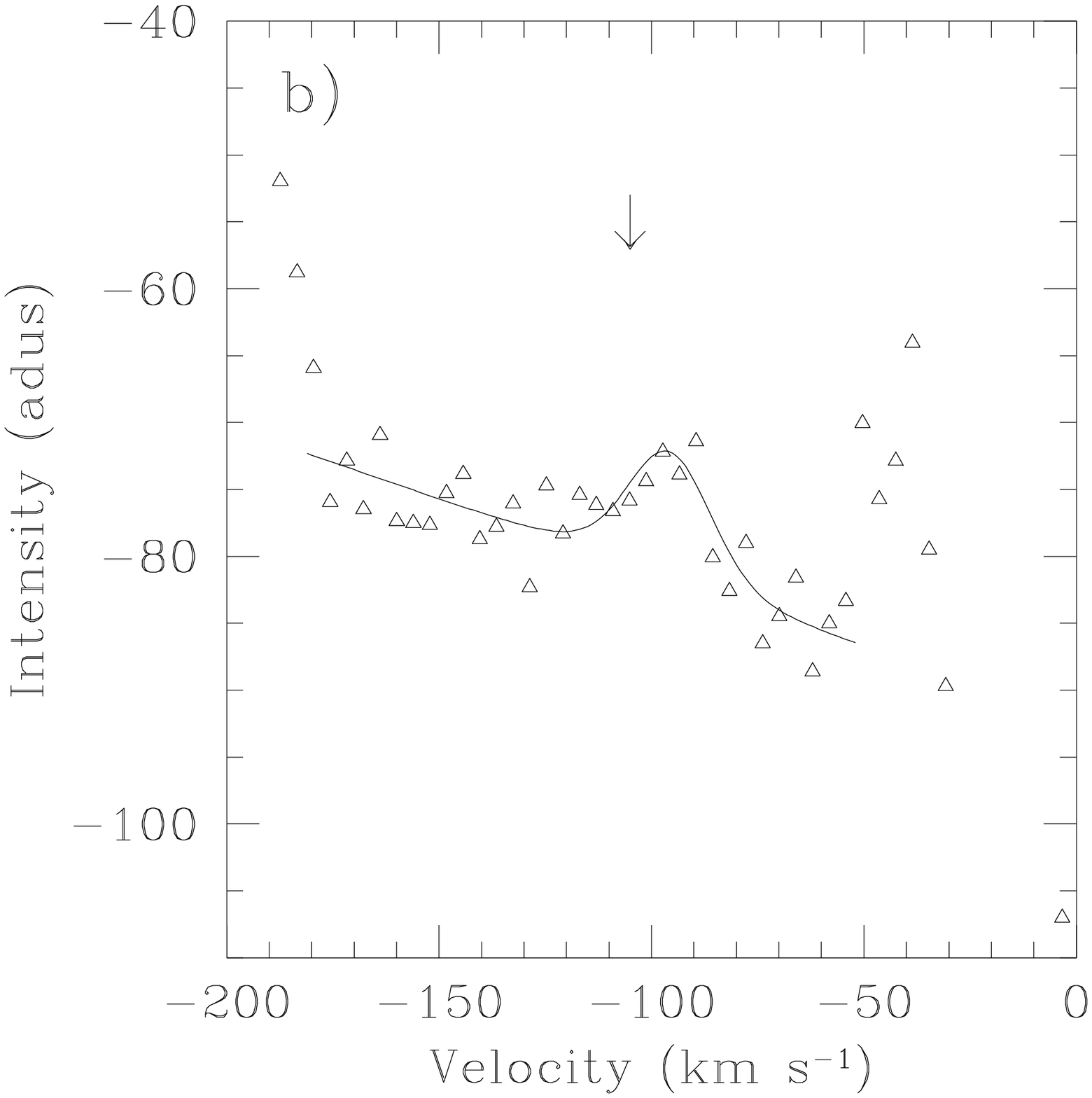}
  \caption{[S~II] from the M~I cloud:  a) 1a - 4a.  b) 6a - 12a (both 
  normalized to 900s exposure).  The arrows show the velocity of the 
  associated H$\alpha$ emission lines.}
\label{fig4-sii_1a4a}
\end{figure}

\begin{figure}[t]
  \centerline{
    \epsfysize = 3.25in
    \epsffile{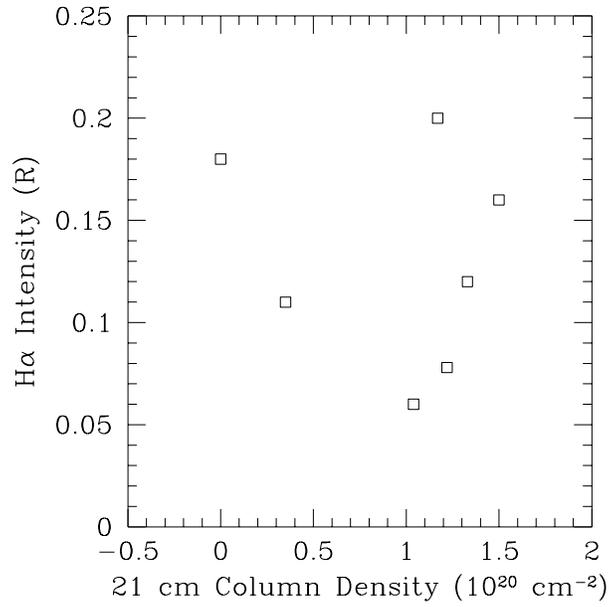}
    }
  \caption{H$\alpha$ intensity versus 21 cm column density for the M Complex}
\label{fig4-inten_corr}
\end{figure}

\begin{figure}[b]
  \centerline{
    \epsfysize = 3.25in
    \epsffile{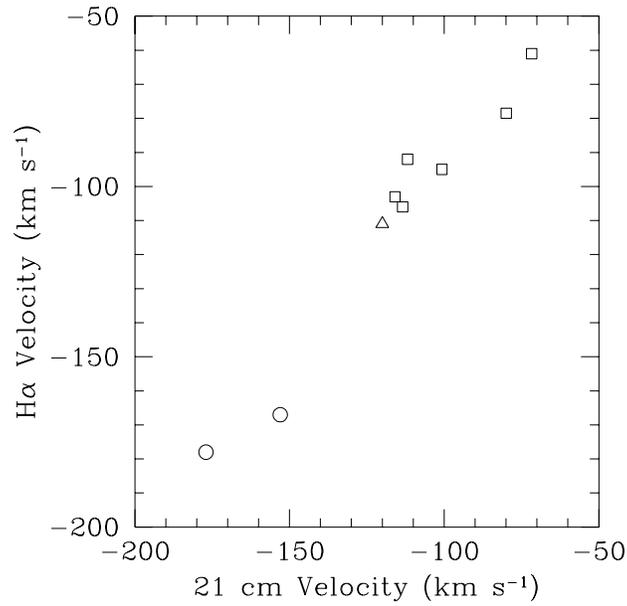}
    }
  \caption{H$\alpha$ velocity versus 21 cm velocity.  Squares: M Complex,
  Circles: A Complex, Triangle: C Complex}
\label{fig4-vel_corr}
\end{figure}

Since interstellar extinction can be neglected along these high
latitude sightlines (especially for the M complex observations), the
H$\alpha$ intensity towards a high-velocity cloud is directly related
to the emission measure (EM $\equiv \int n_{\rm e}^{2} dl$) through
the cloud and gives information on the column density of ionized
hydrogen and the electron density associated with the cloud.  As an
example, consider the 21 cm enhancement centered near direction 1a in
the M complex.  It has an H~I column density of $\simeq$ 1.5 $\times$
10$^{20}$ cm$^{-2}$ (the highest contour level in
Fig.~\ref{fig4-map_m}) and a diameter of about 1.5$^{\rm o}$, which
corresponds to 50 pc at a distance of 2 kpc. Hence the density of
neutral hydrogen n$_{\rm{H~I}}$ $\simeq$ 1.0 cm$^{-3}$, and the mass
of the cloud is about 1600 M$_{\odot}$.  For direction 1a, we measured
I$_{\alpha}$ to be approximately 0.08 R, implying an emission measure
EM = 0.18 cm$^{-6}$ pc, assuming a temperature for the H~II of 8000 K.
This leads to an electron density n$_{\rm e} \simeq$ 0.06 f$^{-1/2}$
cm$^{-3}$ and a column density of ionized hydrogen N$_{\rm{H~II}}
\simeq 1 \times 10^{19}$ f$^{1/2}$ cm$^{-2}$, where f is the filling
fraction of the H~II within the cloud.  The H~I column density in the
1a direction is N$_{H I}$ $\simeq$ 1.22 $\times$ 10$^{20}$ cm$^{-2}$.
Since f $\leq$ 1 and there is little evidence that the H~II associated
with the cloud extends significantly beyond the H~I portion of the
cloud, this would suggest that at this location the HVC is primarily
neutral with N$_{H II}$ / N$_{H I}$ $\lesssim$ 0.08.  Interestingly,
most of the H~II may be located near the outer surface of the cloud.
For example, if we consider direction 6a which appears to intersect
the edge of the cloud (see Fig.~\ref{fig4-map_m}), the H~I column
density is less than 2 $\times$ 10$^{19}$ cm$^{-2}$, whereas the
H$_{\alpha}$ is twice the intensity of 1a, indicating that there may
be as much or more H~II as H~I along the edge of the H~I enhancement.

If the H$\alpha$ emission arises from the photoionization of the
cloud, then the intensity of the emission is directly related to the
incident Lyman continuum flux F$_{LC}$.  Since the H~I cloud is
optically thick in the Lyman continuum and optically thin to H$\alpha$
photons, each Lyman continuum photon incident on the cloud will ionize
a hydrogen atom and each hydrogen recombination will produce on the
average 0.46 H$\alpha$ photons (Martin \markcite{Martin88}1988;
Pengally \markcite{Pengally64}1964; case B, T $\sim$ 10$^{4}$ K).
This leads to the relation
\[ F_{LC}= 2.1 \times 10^{5} \left(\frac{I_{\alpha}}{0.1 R}\right) 
 {\rm \:\:photons\:cm}^{-2} \rm{s}^{-1}. \] 
If, on the other hand, the
ionization is due to a shock arising from the collision of the
high-velocity gas with an ambient medium in the halo, I$_{\alpha}$
provides information on the density and temperature of this ambient
medium.  A cloud velocity of 100 km~s$^{-1}$ is supersonic for ambient
temperatures T $\leq$ 10$^{6}$ K.  If the ambient gas is sufficiently
cool for strong shocks to occur, then we can use shock models by
Raymond \markcite{Raymond79}(1979), which apply to shocks with 50
$\leq$ V$_{\rm s}$ $\leq$ 140 km~s$^{-1}$ and relate the face-on
H$\alpha$ surface brightness (I$_\alpha$)$_\perp$ to the halo gas
parameters. Namely,
\[ (I_{\alpha})_\perp \simeq 6.5 n_{\rm o} 
\left(\frac{V_{\rm s}}{100}\right)^{1.7} R, \] where n$_{\rm o}$ is
the density of the preshocked gas.  This power law is a fit to the
predicted H$\alpha$ intensities for shock velocities between 50
km~s$^{-1}$ and 140 km s$^{-1}$ presented in Raymond
\markcite{Raymond79}(1979).  Hence, for V$_{s}$ = 100 km~s$^{-1}$, and
(I$_{\alpha}$)$_\perp$ = 0.1 R as is the case for the M I cloud, then
n$_{\rm o}$ $\leq$ 0.015 cm$^{-3}$.  This is an upper limit on n$_{\rm
o}$ because a nonperpendicular sightline will increase the observed
I$_\alpha$ for a given n$_{\rm o}$ and V$_{s}$.

The fact that the intensities all seem to be around 0.1 R may be a
clue to the ionization mechanism puzzle.  This fact is more easily
explained in a photoionization scenario in that the various clouds we
measure give independent measurements of the same quantity, that is,
the ionizing flux in the Galactic halo (see Bland-Hawthorn \& Maloney
\markcite{Bland-Hawthorn98}1998).  It is difficult to explain the
uniformity of the H$\alpha$ intensities in a shock excitation
scenario, since in that case the expected H$\alpha$ flux varies
strongly with both the shock velocity and the ambient density.
Another weak piece of evidence for photoionization comes from the
fairly narrow, albeit poorly measured, H$\alpha$ line widths, which
are similar to the widths observed in photoionized H~II regions
surrounding O stars (Reynolds \markcite{Reynolds88}1988).  A more
careful map of I$_{\alpha}$ around the edge of the clouds could also
be used to distinguish between the two ionization mechanisms.

If indeed the detected H$\alpha$ emission results from photoionized
gas, then such measurements can shed light on more general questions
concerning the interstellar medium in the Galactic halo and its
connection to processes occurring in the disk.  For example, consider
the long-standing problem of the ionization of the Warm Ionized Medium
(WIM), an extensive, thick ($\simeq$ 2 kpc) layer of ionized hydrogen
in our Galaxy.  It has frequently been proposed that the source of the
ionization in the WIM is O stars in the plane of the Galaxy
(e.g. Domg\"{o}rgen \& Mathis \markcite{Domgorgen94}1994), although
other, more exotic sources have also been proposed (e.g.  Melott et
al. \markcite{Melott88}1988; Sciama \markcite{Sciama90}1990).  In
order to make the O star idea work, very special arrangements of the
H~I gas must exist to allow photons to get from the Galactic plane,
where the vast majority of O stars are located, to gas high above the
plane in the thick WIM layer (for specific models that have been
constructed to explore this scenario in detail, see Miller \& Cox
\markcite{Miller93}1993 and Dove \& Shull \markcite{Dove94}1994).
However, if the H~I gas is arranged to be very porous to Lyman
Continuum photons, then one would expect a significant fraction of the
photons to leak out of the disk and into the halo.  The H$\alpha$
intensity towards H~I clouds in the Galactic halo constrains the
degree to which this is occurring, and in particular, the H$\alpha$
intensity towards H~I clouds at various distances from the Galactic
plane can reveal the Lyman continuum flux as a function of height
above the plane.  This idea and its implications have been explored
from a theoretical stand point by Bland-Hawthorn \& Maloney
\markcite{Bland-Hawthorn98}(1998).  The H$\alpha$ intensities we measure
for the HVCs are about a factor of 10 lower than typical intensities
from WIM gas at high Galactic latitudes, implying that if the ionizing
photons originate near the Galactic midplane, 90\% of the photons
reaching the WIM are absorbed before reaching the heights of the HVCs
in the Galactic halo.  Note that our measured H$\alpha$ intensity is
an order of magnitude (or more) above what would be expected from the
estimated metagalactic ionizing field flux (Ferrera \& Field
\markcite{Ferarra94}1994).

Perhaps the question of the source of ionization can be further
explored in the future by measuring various emission line ratios.
Models by Raymond \markcite{Raymond79}(1979) predict optical line
ratios for shocks under various conditions.  With an ambient density
n$_{\rm o}$ = 1 cm$^{-3}$ and a shock velocity V$_{s}$ = 100 km
s$^{-1}$, some of the predicted line ratios are [S~II] $\lambda$6716 /
H$\alpha$ = 0.18, [N~II] $\lambda$6584+$\lambda6548$ / H$\alpha$ =
0.51, and [O~III] $\lambda$5007+$\lambda$4959 / H$\alpha$ = 1.9.  If
this model, when extended to lower densities, retains the prediction
of very bright [O~III] emission, then the $\lambda$5007 line could be
an important discriminator between shock ionization and
photoionization.  Note that these emission lines can also give
further information on the metal abundances in the HVCs, which have
been previously investigated by Lu et al. \markcite{Lu94}1994 and Lu
et al. \markcite{Lu98}1998, through absorption line techniques.

It should be noted that the physical conditions in the Galactic halo
may vary greatly from location to location.  For example, the
absorption line measurements of highly ionized halo gas towards
Markarian 509 (Sembach et al. \markcite{Sembach95}1995, Sembach et
al. \markcite{Sembach98}1998) suggest a hot ($\sim$ 10$^{5}$ K) halo
around this HVC.  Also, an analysis of Rosat observations of soft
X-rays combined with 21 cm data from the new Leiden/Dwingeloo survey
toward complex M by Herbstmeier et al.\ \markcite{Herbstmeier95}(1995)
suggests that there is a brightening of soft X-rays at several cloud
locations, perhaps due to collisions of the clouds with the ambient
medium in the halo.  The most striking such enhancement occurs on the
edge of cloud M~I, where the H~I contours are also seen to be crowded.
It should be noted, however, that considerable modeling is necessary
to separate out the soft X-rays possibly associated with the M complex
region from sources of soft X-rays at other distances in the same
direction.  The pattern of H$\alpha$ intensities, tracing the 10$^{4}$
K ionized gas, does not seem to correlate well with the soft X-ray
brightness derived by Herbstmeier et al.
\markcite{Herbstmeier95}(1995).  They calculate that there is an
enhancement in soft X-rays emitted by the M~I cloud in direction 1a.
We do not see enhanced H$\alpha$ in this direction.  Direction 5a,
with no strong X-ray enhancement on the Herbstmeier et al. map, has
both comparable 21 cm emission and comparable H$\alpha$ emission (see
Table~\ref{tab4-obs_results}).  On the M~II cloud, the 1b direction is
at the location of another of the Herbstmeier et al.\ soft X-ray
enhancements, and the 2b direction is located where the soft X-ray
emission is thought to be faint.  We, on the other hand, find the 2b
direction to be significantly brighter in H$\alpha$ than the 1b
direction.

Finally, improvements in the observational technique may help to
eliminate possible systematic errors in line widths and velocities.
For example, it is clear from the results in
Section~\ref{sec4-results} that it would be extremely valuable to find
an ``off'' direction with zero Galactic H$\alpha$ emission at all
velocities.  This would allow one to subtract the sky spectrum and
correct for other systematic uncertainties without introducing the
uncertainty caused by an unknown quantity of emission from the ``off''
direction.  The Lockman Window direction (our 10a) appears to be a
good approximation to this ideal.  This direction has a famously low
H~I column density (N$_{H I}$ = 4.4 $\pm$ 0.5 $\times$ 10$^{19}$
cm$^{-2}$; Jahoda, Lockman, \& McCammon \markcite{Jahoda90}1990), and
as a result is often used by X-ray astronomers and others interested
in looking at sources unobscured by the affects of neutral hydrogen
and the associated dust.  It also appears to be fainter in H$\alpha$
than any other direction we have observed, with H$\alpha$ components
no brighter than 0.1 R\@.  We are at present carefully examining the
spectrum in this direction to determine its usefulness as a ``zero
emission'' standard.  We will eventually be able either to detect
H$\alpha$ or put very low upper limits on its intensity by using the
shifts in velocity caused by the earth's orbital motion to distinguish
between any actual emission and baseline irregularities or night sky
features.  If this direction can be used in this way it will not only
improve H$\alpha$ studies of the HVC complexes, but also will allow
the study of IVCs in H$\alpha$.

We thank Nicole Hausen for her work on the weak atmospheric lines.  This 
work was supported by the National Science Foundation through grants
AST 9122701 and AST 9619424.

\clearpage


\begin{deluxetable}{cllcc} 
\tablecolumns{5}
\tablewidth{0pt}
\tablecaption{M Complex H$\alpha$ Observation Data.}
\tablehead{
\colhead{Name} & \colhead{$l$ (\arcdeg)}   & \colhead{$b$ (\arcdeg)}   
& \colhead{Exp. Time} & \colhead{Obs. Date \tablenotemark{a}} \\
\colhead{} & \colhead{} & \colhead{} & \colhead{(seconds)} & \colhead{}
}
\startdata
1a   & 168.30       & 65.20        & 4500          & 1,2,3 \nl
2a   & 163.30       & 66.70        & 1800          & 2     \nl
5a   & 165.80       & 65.95        & 900           & 3     \nl
6a   & 170.85       & 64.72        & 900           & 4     \nl
7a   & 168.30       & 67.82        & 900           & 3     \nl
8a   & 163.30       & 64.70        & 900           & 3     \nl
&&&& \nl
1b   & 183.40       & 65.75        & 900           & 2     \nl
2b   & 186.00       & 65.00        & 900           & 2     \nl
&&&& \nl
4a   & 168.00       & 70.93        & 4500          & 1,2,3 \nl
3a   & 168.30       & 69.00        & 1800          & 2     \nl
9a   & 167.00       & 60.00        & 900           & 3     \nl
12a  & 169.00       & 74.00        & 900           & 4     \nl
11a  & 140.00       & 80.00        & 900           & 3     \nl
10a  & 150.48       & 52.96        & 900           & 3     \nl
&&&& \nl
3b   & 186.0        & 70.00        & 900           & 2     \nl
\enddata
\tablenotetext{a}{Observations were carried out on 4 nights, \\ 
1: 2/18/96,\ \ 2: 2/24/96,\ \ 3: 3/23/96,\ \ 4: 4/10/96.}
\label{tab4-obs_info}
\end{deluxetable}

\clearpage

\begin{deluxetable}{rrrrrccc} 
\tablecolumns{8}
\tablewidth{0pt}
\tablecaption{H$\alpha$ and 21 cm Emission from HVCs}
\tablehead{
\colhead{}  &  \multicolumn{3}{c}{H$\alpha$}  &  \colhead{}  &
\multicolumn{3}{c}{21 cm} \\
\cline{2-4} \cline{6-8} \\
\colhead{Name} & \colhead{I$_{H\alpha}$} & \colhead{V$_{H\alpha}$} &
\colhead{W$_{H\alpha}$} & \colhead{} & \colhead{I$_{21cm}$} & 
\colhead{V$_{21cm}$} & \colhead{W$_{21cm}$} \\
\colhead{} & \colhead{(R)} & \colhead{(km s$^{-1}$)} &
\colhead{(km s$^{-1}$)} & \colhead{} & \colhead{(10$^{20}$ cm$^{-2}$)} & 
\colhead{(km s$^{-1}$)} & \colhead{(km s$^{-1}$)} }
\startdata
M I cloud & & & & & & & \nl 
1a-4a  & 0.08 $\pm$ 0.01 & -106 $\pm$ 1 & 27 $\pm$ 4  & & 1.22 & -113 & 24 \nl
2a-3a  & 0.15 $\pm$ 0.02 & -98  $\pm$ 1 & 38 $\pm$ 3  & & 1.12 & -102 & 40 \nl
2a-4a  & 0.20 $\pm$ 0.02 & -95  $\pm$ 1 & 49 $\pm$ 4  & & 1.17 & -101 & 43 \nl
5a-9a  & 0.09 $\pm$ 0.02 & -109 $\pm$ 2 & 46 $\pm$ 8  & & 1.08 & -116 & 31 \nl
5a-4a  & 0.06 $\pm$ 0.02 & -103 $\pm$ 3 & 38 $\pm$ 10 & & 1.04 & -116 & 30 \nl
6a-12a & 0.18 $\pm$ 0.02 & -105 $\pm$ 1 & 35 $\pm$ 4  & & \nodata & \nodata &
\nodata \nl
8a-4a  & 0.11 $\pm$ 0.02 & -92  $\pm$ 2 & 50 $\pm$ 7  & & 0.35 & -112 & 26 \nl
 & & & & & & & \nl 
M II cloud & & & & & & & \nl 
1b-4a  & 0.12 $\pm$ 0.01 & -61  $\pm$ 2 & 44 $\pm$ 5  & & 1.33 & -72  & 22 \nl
2b-3b  & 0.20 $\pm$ 0.02 & -72  $\pm$ 1 & 44 $\pm$ 4  & & 1.49 & -80  & 26 \nl 
2b-4a  & 0.16 $\pm$ 0.03 & -78  $\pm$ 2 & 27 $\pm$ 5  & & 1.50 & -80  & 25 \nl 
 & & & & & & & \nl 
A cloud & & & & & & & \nl 
A III  & 0.08 $\pm$ 0.01 & -167 $\pm$ 1 & 24 $\pm$ 2  & & 1.40 & -153 & 25 \nl
A IV   & 0.09 $\pm$ 0.01 & -178 $\pm$ 1 & 34 $\pm$ 3  & & 1.34 & -177 & 23 \nl
 & & & & & & & \nl 
C cloud & & & & & & & \nl 
C      & 0.13 $\pm$ 0.03 & -111 $\pm$ 2 & 38 $\pm$ 7  & & 0.54 & -120 & 15 \nl
\enddata
\label{tab4-obs_results}
\end{deluxetable}

\clearpage


\clearpage

\clearpage

\clearpage

\clearpage

\clearpage

\end{document}